\documentclass[notoc,12pt]{JHEP3} 
\usepackage{amsmath}
\input epsf.tex
\usepackage{amssymb,amsfonts}

\def\be{\begin{equation}}
\def\ee{\end{equation}}
\def\ba{\begin{eqnarray}}
\def\ea{\end{eqnarray}}

\def\sqr#1#2{{
\vcenter{\vbox{\hrule height.#2pt
\hbox{\vrule width.#2pt height#1pt \kern#1pt
\vrule width.#2pt}
\hrule height.#2pt}}}}
\boldmath
\title{Permeable conformal  walls  and holography}
\unboldmath
\author{C. Bachas$\;^{1,5}$, J. de Boer$\;^{2,5}$, 
R. Dijkgraaf$\;^{2,3,5}$
 and H. Ooguri$\;^{4,5}$
\\
$\ $ \\
$\ $ \\
$\;^1$Laboratoire de Physique Th{\'e}orique
de l'Ecole Normale Sup{\'e}rieure\thanks{Unit{\'e} mixte  du
CNRS et de l'Ecole Normale Sup{\'e}rieure,
UMR 8549.} \\ 
$\;\;\;$24  rue Lhomond, 75231 Paris Cedex 05, France\\ 
$\ $ \\
$\;^2$Institute for Theoretical Physics, University of Amsterdam\\
$\;\;\;$Valckenierstraat 65, 1018 XE Amsterdam, The Netherlands\\
$\ $ \\
$\;^3$Korteweg-de Vries Institute for Mathematics, 
University of Amsterdam\\
$\;\;\;$Plantage Muidergracht 24, 1018 TV Amsterdam, 
The Netherlands\\ 
$\ $ \\
$\;^4$California Institute of Technology 452-48, 
Pasadena, CA 91125, USA\\
$\ $ \\
$\;^5$Institute for Theoretical Physics, University of California\\
$\;\;\;$Santa Barbara, CA 93106-4030, USA
\\
}

\abstract{We  study 
conformal field theories in two dimensions 
separated by domain walls,  
which preserve  at least one Virasoro algebra.
We develop tools to study such domain walls, extending
and clarifying the concept of  `folding'
discussed in the condensed-matter literature. We analyze the
conditions for unbroken supersymmetry,  
and discuss  the  holographic duals  in AdS3  when
they exist. One of the interesting observables is
the Casimir energy between a  wall and an  anti-wall.
When these  separate free scalar field theories
 with different target-space radii, the Casimir
energy is given by the dilogarithm function
of the reflection probability. The walls with
holographic duals  in AdS3 separate  two sigma models,  whose target spaces
are moduli spaces of Yang-Mills instantons
on T4  or K3.  In the supergravity
limit, the Casimir energy is computable as
classical energy of  a brane that connects
the  walls through AdS3. We compare this 
result with expectations from the sigma-model
point of view.
}

\preprint{
CALT-68-2361\\ 
CITUSC/01-045\\
ITFA-2001-33\\ 
LPTENS-01/42
\\hep-th/0111210}

\begin{document}

\section{Introduction}

  Starting with  the pioneering work  of Cardy \cite{Ca},  
boundary conformal field theory (BCFT) has evolved into a rich subject
of great physical interest. The subject is of obvious relevance to 
the study of  critical phenomena in statistical mechanics. 
Furthermore,  two-dimensional conformal boundary states   
have acquired  new importance in recent years,  as building blocks for
the D(irichlet) branes of string theory
\cite{Po}.  The interplay between the  algebraic  approach
of Conformal Field Theory,  and the
complementary geometric viewpoint of  
D-branes, has been   the theme  of many
recent investigations
 (see e.g.  \cite{rev1,rev2} and references therein).

  The usual setting of BCFT is a space(time) { ending}  on a  boundary.
In this  setting  all incident waves 
are  reflected back,\footnote{The language is somewhat loose, because
strictly-speaking a  CFT has no
  asymptotic particle states. A more accurate phrasing,  in two
dimensions, is that the boundary  state maps  holomorphic into
antiholomorphic fields, in a way that commutes with  the action of the Virasoro
algebra.}
because  there is nothing they can transmit to on the other side. 
One may, however, also consider a  situation in which  
two (or more) non-trivial CFT's are glued together along a common interface. 
The interface  can be  {permeable}, meaning that  incident waves
are  partly  reflected  and  partly transmitted.
Examples of such boundaries (mostly between identical CFT's)
have been discussed in the
condensed-matter literature, see for instance \cite{OA,LL,LL1,AOS}. 
One  of our  purposes in this  work will be  to analyze such 
permeable  interfaces in  general, and  from a rather  different, 
more geometric perspective.

   Our interest in  these  questions  was motivated by an issue 
in  holography.
String theory in AdS3 
has static solutions describing   infinitely-long
  $(p,q)$  strings, which   
stretch  between two  points on the AdS3 boundary \cite{BP}.
In the dual spacetime CFT \cite{MS,Ma,De,GKS,Di,BORT}
the endpoint  of a  $(p,q)$   string is, as we
will explain, an  
interface separating regions  with different values of
the  central charge or different values of the moduli. 
Similar configurations
have been  also discussed in higher dimensions \cite{KR,DeWolfe}.
The force exerted by 
the stretched  string on its endpoints  translates,  in 
the  dual interpretation,  to the  Casimir force
 between two (or more, if one considers
 string networks)  permeable interfaces.  
In this paper we  will calculate
this  Casimir  force,  
 both in the weak- and in the strong-coupling limits. 
The results we find are 
in some ways reminiscent of the heavy quark--antiquark
potential in  four-dimensional N=4 super Yang--Mills
\cite{qq1,qq2}.

  From a technical point of view,  an
 interface between two CFTs  is described by a 
regular boundary state in
  the tensor-product theory.\footnote{More precisely,  the
tensor product of the theory on one side  and of  the `conjugate'  theory, 
with  left-and right-movers interchanged, on the other side.}
This is  intuitively obvious 
since one can  `fold' space along the interface, so that both CFTs live
 on the same side \cite{OA}. 
Permeable  walls, in particular, are simply 
  boundary states of the tensor product, 
that cannot be expressed in terms of Ishibashi states of the
factor theories. 
Their study does not,  therefore,   require
drastically-new technology,  but it leads  to  
a host of novel  questions and  observables which  are
not usually considered in the standard BCFT setting. 
One example of such a new observable  is the
Casimir energy of a `CFT bubble' which  we calculate.

The plan of this paper is as follows. In section 2 we 
 introduce the main ideas of `conformal gluing'   in the simplest  context 
of a free scalar field theory, and  explain how  this is related 
to conventional  conformal boundary states. We  calculate
the Casimir energy for  two identical interfaces, separating regions
with different target-circle  radii,   
and show that it
is given by  the dilogarithm function
of the reflection probability. 
 In section 3 we generalize these
considerations in several directions. We show  how superconformal
invariance of the walls can be   guaranteed by the continuity of
 appropriately-defined `half' superfields, in a manifestly
 supersymmetric formalism. We also   calculate the fermionic
 contribution to the Casimir energy, and then go on to  discuss general
 properties of permeable interfaces  and some more examples.
 In section 4 we turn our
 attention to interfaces  of  two-dimensional  CFTs which admit 
 holographic AdS3 duals. We calculate the classical energy of a
 $(p,q)$ string  as a function of its tension, Neveu--Schwarz-charge and 
 of  the separation of its
 two endpoints. We  discuss the validity of this  calculation, 
and  interpret it  as  Casimir
 energy in  the dual spacetime
sigma model. We  point out an intriguing analogy 
with operator algebras
on instanton moduli spaces defined in the mathematics
 literature \cite{Grojnowksi,Nakajima1,Nakajima2}. 
We conclude,  in section 5,  with some comments on
future directions.

\vskip 0.4cm


\section{Free scalar field}
 
 In this section we  discuss  conformal `permeable' walls
for a single free scalar field $\phi$. 
This is the simplest setting in which to illustrate the main ideas
and calculation tricks, which we will then  apply and extend 
to other contexts.


\subsection{Gluing conditions}

Consider a free massless scalar field in 1+1 dimensions,  $\phi(x,t)$.
We are interested in  scale-invariant defects  
described by the  `gluing'   conditions:
\begin{equation}
\left(
\begin{array}{l}
\partial_x \phi\cr
\partial_t \phi\cr
\end{array}
\right)_{x=-0}
=\;\;
M\; 
\left(
\begin{array}{l}
\partial_x \phi\cr
\partial_t \phi\cr
\end{array}
\right)_{x=+0}
\label{bc}
\end{equation}
where $\pm 0$ denote points just
to  the left or  right  of the wall, which is located at $x=0$, 
and  $M$ is a constant  $2\times 2$ matrix.
Energy conservation requires that\footnote{The light-cone coordinates are 
taken to be $x^\pm  = t \pm x$,
so that $\partial_\pm = {1\over 2}(\partial_t \pm \partial_x)$.} 
\begin{equation}
T_{xt}\; = \;
  T_{++}-  T_{--} \;=\;
 \partial_x\phi\; \partial_t \phi
\label{conf}
\end{equation}
be continuous across the defect. Alternatively, notice that  the conformal
transformations which leave invariant the $x=0$ worldline, 
are generated by the operators 
$[f(x^+)T_{++}-f(x^-)T_{--}]$ . In the Wick-rotated  theory, we can obtain
the corresponding 
Ward identities  by 
inserting a contour integral
of these  operators  in correlation functions.
Continuity of \eqref{conf} ensures that one can  deform
the contour,  so as to only pick contributions  from
field insertions. This is illustrated in figure 1.

\begin{figure}[ht]
\hskip 3.9  cm
       \hbox{\epsfxsize=78mm%
       \hfill~
       \epsfbox{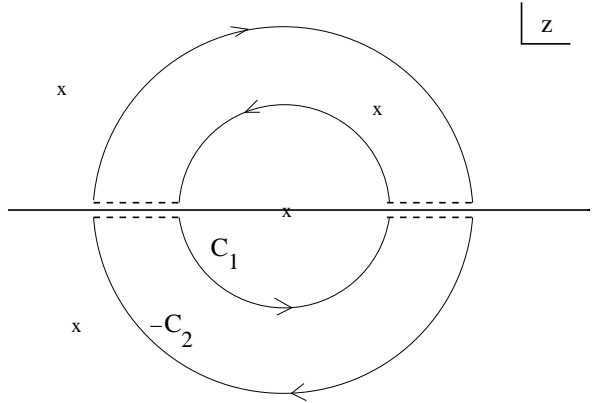}
       \hfill~}
       \caption{Conformal Ward identities are obtained  by inserting 
\  $\ \oint_C [T f(z)dz - 
{\bar T}{ f}(\bar z)d{\bar z}]$  in correlation functions. 
In deforming  the contour from $C_1$ to $C_2$ we pick up
contributions from  the broken-line segments. These cancel out provided 
$T-{\bar T}$  is continuous. 
The crosses in the figure stand for local field insertions. 
}
\end{figure}

The continuity of $T_{xt}$ implies
 that $M$ must be an element of O(1,1) . 
 This  group has four  disconnected  components,
\begin{equation}
M  =  \pm  \; \left(
\begin{array}{lll}
\lambda & \quad &  0\cr
\; 0 & \quad & \lambda^{-1}\cr
\end{array}
\right) \  \ \ 
\ \ \ {\rm or} \ \ \ \ \ M^\prime  =  \pm \; \left(
\begin{array}{lll}
\; 0  & \quad &  \lambda^{-1} \cr
\lambda & \quad &  0 \cr
\end{array}
\right)\ , 
\label{O11}
\end{equation}
with  
 $\lambda$ a real positive number. 
We will  `compactify'  the group  by allowing also the 
singular  values $0$   and $\pm \infty$, so that $\lambda$ runs
over the entire compactified real line.
 As a result, the four disconnected
components merge into two, which can be  parametrized as follows:
$$ 
 M(\vartheta)\ \ \ {\rm  and}\ \ \  
M^\prime(\vartheta)\  , \ \ \  {\rm with} \ \ \ 
\vartheta\; \equiv \; {\rm arctan}\lambda\;
\in\; [-\pi/2, \pi/2]\ .
$$
We will  see in the following subsections
 that  this parametrization is natural.

The singular  values of $\lambda$ 
correspond  to 
{\it perfectly reflecting}  defects, for which  the fields
on either  side  don't communicate.
Gluing  derivatives with $M(0)$,  for example, implies that
$\partial_t\phi(+0) = 
\partial_x\phi(-0) = 0$. This  is a  standard Neumann  condition
for the field to  the
left of the wall, and a Dirichlet  condition for the field on   the right.
Let us denote it  
  by  `ND' (not to be confused with the mixed boundary  conditions
one often  writes for  the annulus).
As  can be,  likewise,  easily checked,  
$M(\pm \pi/2)$,  
$M^\prime(0)$ and $M^\prime(\pm \pi/2)$
correspond, respectively,  to DN,  NN  and DD  boundary conditions.

\begin{figure}[ht]
 \hskip 3.0  cm
       \hbox{\epsfxsize=95mm%
       \hfill~
       \epsfbox{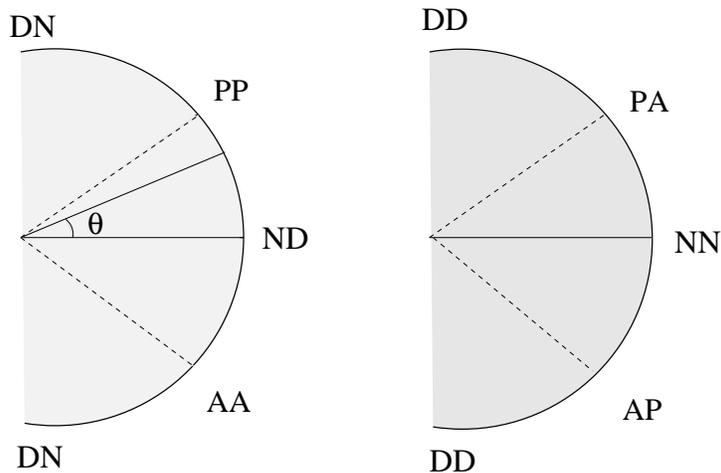}
       \hfill~}
       \caption{The moduli space  of
gluing matrices,   $M(\vartheta)$ on the left and
  $M^\prime(\vartheta)$ on the right, where 
$\vartheta = {\rm arctan}\lambda \in [-\pi/2,\pi/2]$.
  Perfectly-reflecting walls
are labeled by the two boundary conditions, 
Dirichlet (D) or Neumann (N),  on either side of the defect.
Totally-transmitting defects are labeled by the periodicity
properties of $(\partial_-\phi,\partial_+\phi)$ when
$x$ is compactified on  a circle. 
 }
\end{figure}

At the opposite extreme of the spectrum one has the four
{\it perfectly  transmitting} cases, corresponding to  the special values
$\vert \lambda\vert  =1$. Clearly, 
 $M(\pi/4) =  {\bf 1}$ gives  continuous  derivatives -- there
is no defect in this special case.  Gluing with 
$M(-\pi/4)$ makes  $\phi$ jump  to $-\phi$, but both left- and right-moving
waves are still fully transmitted.
 The same is true for the two `chiral defects' 
$M^\prime(\pm \pi/4)$. For one of them    left-moving  waves are
 continuous across the wall,
while right-moving waves   pick a minus
 sign. For the other, the roles of left
and right are reversed. If we were to let  $x$ be an angle coordinate, the
 four  perfectly-transmitting
walls would give rise to 
PP, AA, PA and AP boundary conditions for 
($\partial_-\phi$, $\partial_+\phi$).

 The  general defects  interpolate  between these standard 
cases. They are `permeable', i.e.
 partially-reflecting and partially-transmitting.
The two disconnected components
of their moduli space are exhibited  as two half-circles in figure 2. 
Sending $\lambda\to 1/\lambda$  
exchanges, as can be easily seen, $x$- and $t$-derivatives
on both sides.
This is, therefore,   the action of a  T-duality transformation
on the `permeable  defects.


\subsection{S-matrix and Casimir energy}

 The defects in the first connected component of O(1,1)  have  a
simple realization as  discontinuities 
 in  the radius of compactification
of the scalar field. Indeed,  let the
field ${\tilde\phi}
\equiv  {\tilde\phi}+2\pi$  
be  continuous in the entire plane,   but have 
a  discontinuous
action
\begin{equation}
{ I}   = 2 r_1^{2} \int_{x < 0} 
 \partial_+\tilde\phi\; \partial_-\tilde\phi\  +\ 
2 r_2^{2} \int_{x>0}  \partial_+\tilde\phi\; \partial_-\tilde\phi\ .
\end{equation}
Varying $I$  gives the boundary  conditions at $x=0$\;:
\begin{equation}
r_1^{2}\;\partial_x \tilde\phi \; \Bigl\vert_{-0}
 \; =\;  r_2^{2}\;\partial_x \tilde\phi \;
 \Bigl\vert_{+0}\ \ .
\end{equation}
Redefining   the scalar field so as to normalize its energy-momentum
tensor,
$$
 \phi \equiv  \Biggl\{ 
\begin{array}{lll}
r_1 \tilde\phi   & \quad & x < 0  \cr
r_2 \tilde\phi   & \quad & x >0  \  ,  \cr
\end{array}\ 
$$
leads   precisely to the discontinuity equation \eqref{bc}, 
where the argument of the gluing matrix $M(\vartheta)$ obeys
\begin{equation}
{\rm tan}\;\vartheta \; = \;\lambda\; =\; {r_2\over r_1}\ \; .
\label{theta}
\end{equation}
Thus, the parameter
 $\lambda= {\rm tan}\;\vartheta $ is  related to the multiplicative
discontinuity of the compactification radius across the wall.
We will see the geometric significance of this
 fact in the following subsection.

\vskip 0.3cm

\begin{figure}[ht]
 \hskip 3.0  cm
       \hbox{\epsfxsize=65mm%
       \hfill~
       \epsfbox{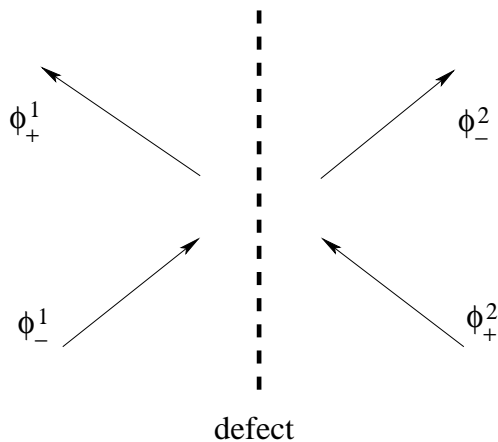}
       \hfill~}
       \caption{The incoming and outgoing waves can be related by the matrix
       $S$ .}
\end{figure}

 Another  useful  characterization of  the defects  is
 in terms of a `scattering matrix', 
from which one can read directly the
  reflection and transmission coefficients.
 Let us, for ease of notation, call  $\phi^1$ the
field to  the left of the wall, and $\phi^2$ the field to 
the right. Then $\partial_- \phi^1$ and $\partial_+ \phi^2$ can
be expanded in terms of
 `incoming waves', while  $\partial_+ \phi^1$ and $\partial_- \phi^2$
can be expanded in terms of `outgoing waves' (as
illustrated in  figure 3). Strictly-speaking
one cannot define  asymptotic states for a massless 2d  field,
but this will  not be important  for our discussion  here.

 With the help of some linear algebra,
we  can write  the gluing conditions \eqref{bc} in the equivalent form
\begin{equation}
\left(
\begin{array}{l}
\partial_- \phi^1\cr
\partial_+ \phi^2\cr
\end{array}
\right)
=\;\;
S \; 
\left(
\begin{array}{l}
\partial_+ \phi^1\cr
\partial_- \phi^2\cr
\end{array}
\right) \ \ , 
\label{bcS}
\end{equation}
where
\begin{equation}
S  =   \; \left(
\begin{array}{lll}
  -{\rm cos}\;2\vartheta & \quad &  {\rm sin}\;2\vartheta \cr
 \;\;{\rm sin}\;2\vartheta & \quad & {\rm cos}\;2\vartheta \cr
\end{array} 
\right) \  \ \ {\rm and}\ \ \ 
 S^\prime  =   \;  \left(
\begin{array}{lll}
\;{\rm cos}\;2\vartheta  & \quad & -{\rm sin}\;2\vartheta \cr
 \;{\rm sin}\;2\vartheta & \quad &  \;\;{\rm cos}\;2\vartheta \cr
\end{array}
\right)\ .  
\label{SO11}
\end{equation} 
The orthogonal matrices  $S$  and $S^\prime$
relate  incoming to  outgoing waves at the defect. They are
independent of  the wave-frequency, as required by conformal invariance. 
 Furthermore, they are  off-diagonal
for $\vartheta=\pm \pi/4$, corresponding to a perfectly-transmitting defect, 
and diagonal  for $\vartheta$ a
multiple of $\pi/2$, which corresponds to total reflection (see figure 2).

  One  simple  observable,  that can be expressed in terms
of the scattering matrix,  is   the
Casimir force   between a  defect  and an anti-defect.
Consider, to be specific,  an interval
 inside which the radius of 
the  scalar field jumps from $r_1$ to $r_2$, 
\begin{equation}
I  = 
\left( 2r_1^{2} \int_{-\infty}^{-d} \ \; +\ \;
2r_2^{2} \int_{- d}^{\; d} \ \; +\ \;  2r_1^{2} \int_{d}^{\;\infty}\right)
  \partial_-\tilde\phi\;  \partial_+\tilde\phi\ \ .
\end{equation}
We assume that $\tilde\phi$ is continuous in the entire plane. 
It follows from our previous discussion,
that there  is   a defect 
 $M(\vartheta)$ located at $x=-d$,  and an anti-defect 
with gluing matrix $M(\pi/2- \vartheta)$ at  $x=d$, where 
  $\vartheta$ is given by  equation \eqref{theta}. 
The setup  is illustrated in figure 4.

\vskip 0.3cm

\begin{figure}[ht]
 \hskip 3.0  cm
       \hbox{\epsfxsize=85mm%
       \hfill~
       \epsfbox{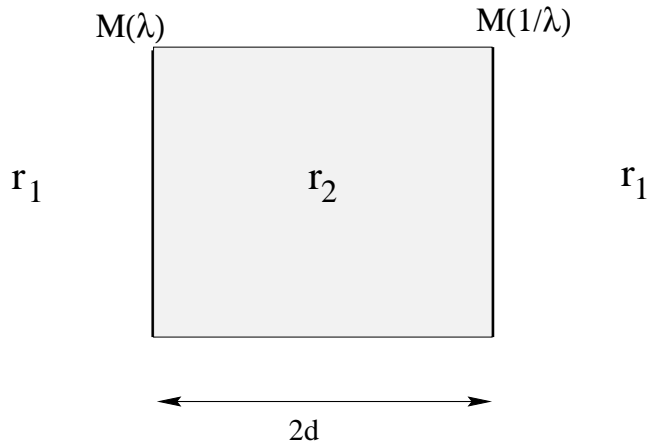}
       \hfill~}
       \caption{The  region  of rescaled radius ($r_2=\lambda r_1$)
  bounded by a defect and
an anti-defect. Time runs in the upward direction.
The interfaces feel  an  attractive Casimir  force. }
\end{figure}

In order to  calculate  the zero-point energy, we put  the
configuration in a larger box of size $2L$ so as to discretize the 
allowed frequencies. The presence of the defects in the middle induces
a $d$-dependent shift in these frequencies, thereby modifying the
zero-point sum. Taking $L\to\infty$ removes the dependence on
the precise boundary conditions in the larger box, which can thus
be chosen at will for convenience. What is left behind is a Casimir
energy describing the
interaction of the wall and antiwall. The calculation is rather subtle,
because
of the need to regularize the ultraviolet, and  can
be found  in appendix A. The result is 
\begin{equation}
{\cal E} =  - \frac{1}{8\pi d }\;  {\rm Li}_2\;({\cal R}^2)\ \ 
\ ,
\label{dilog} 
\end{equation}
where ${\rm Li}_2(x) = \sum_1^{\infty} x^n/n^2$  is the dilogarithm
 function \cite{lewin},
 and 
 ${\cal R}$ is the reflection amplitude, 
\begin{equation}
{\cal R}= {\rm cos}\; 2\vartheta = {1-\lambda^2\over 1+\lambda^2}\ . 
\end{equation}
For weak reflection the energy
vanishes (as it should) quadratically: 
\begin{equation}
{\cal E} \simeq  - {{\cal R}^2\over 8\pi d } + o({\cal R}^4)\; .  
\end{equation}
Total reflection, on the other hand, corresponds to 
${\cal R}=\pm 1$. Since   ${\rm Li}_2(1) =  \pi^2/6$, 
one recovers  the standard Casimir energy for a massless scalar field
in a box in this special  case.


\boldmath
\subsection{Folding trick}
\unboldmath

 The permeable defects of the previous sections can be described
as regular D-branes, after  `folding' the plane along the defect line.
This simple but powerful trick is well-known in the condensed-matter
literature, and has been used for instance in 
the  study of fracture lines for the Ising model \cite{OA}.
To be more precise, let us  define a `conjugate' 
 field in the left-half plane by
mirror reflecting the field on the right, 
\begin{equation}
{\hat \phi}^2(x,t) \equiv \phi^2(-x,t)\ \ \ {\rm for}
\ \ \  x\le 0 .
\end{equation}
The  gluing  conditions \eqref{bc} with gluing matrix $M(\vartheta)$ read: 
\begin{equation}
\partial_t\;({\rm cos}\;\vartheta\; \phi^1 - {\rm sin}\;\vartheta\;
{\hat\phi}^2)\Bigl\vert_0 \; =\;
\partial_x\;({\rm sin}\;\vartheta\; \phi^1 +  {\rm cos}\;\vartheta\;
{\hat\phi}^2)\Bigl\vert_0 \; = \;  0\ .
\label{D1}
\end{equation}
These are the  boundary conditions for  a D1-brane stretching along the
direction $\vartheta$ in the $(\phi^1,{\hat\phi}^2)$ plane.
The parametrization  of the defects 
 in terms of an angle variable  can now be recognized as
most  natural. Note that
bosonic D-branes are unoriented, which is why $\vartheta$ 
 runs only  over half a circle.
Note also that  the
 relation \eqref{theta} between $\vartheta$ and the radii, in the case of 
periodically-identified fields, ensures that the D1-brane is compact. 
These facts are illustrated in figure 5.

\begin{figure}[ht]
 \hskip 3.0  cm
       \hbox{\epsfxsize=95mm%
       \hfill~
       \epsfbox{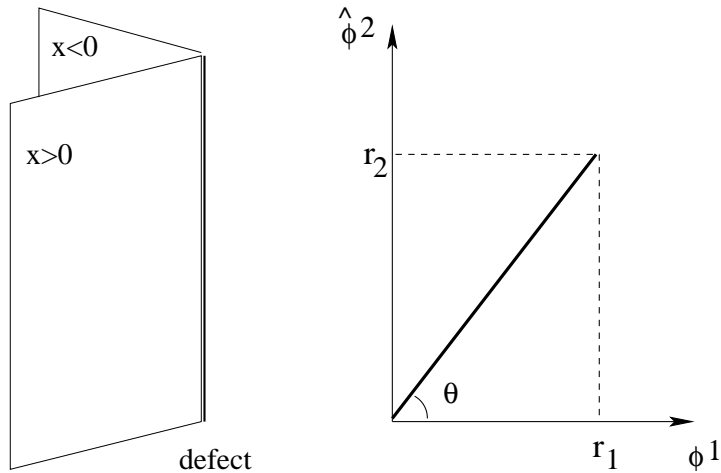}
       \hfill~}
       \caption{Folding the plane along the defect line leads to a 
description of the permeable defects as regular D-branes in a two-dimensional 
target space.}
\end{figure}

 To see  the power of the  folding trick, let us now rederive
the Casimir energy of the previous subsection. 
We will need  the conformal boundary
 state (see  \cite{paolo,ga} for  nice reviews)
that describes  the D1-brane \eqref{D1}
in the closed-string language, 
\begin{equation}
\vert \vartheta \gg \  =\  {\cal N}\;
\prod_{n=1}^\infty {\rm exp}\left(
{1\over n} a^i_{-n} {\tilde a^j}_{-n} S_{ij}
 \right)\;  \vert 0; \varphi_\perp, w_\parallel> \
.
\label{bstate}  
\end{equation}
Here $a^{1,2}_n$ are the canonically-normalized
left-moving oscillators for the 
fields $\phi^1$ and $\hat\phi^2$, and
 ${\tilde a}^{1,2}_n$ are the corresponding right-moving oscillators.
Note that these are the oscillators in the closed-string channel, where
the roles of space and time are interchanged. 
The matrix $S$ is  given by equation  \eqref{SO11}, and ${\cal N}$ is 
a normalization factor. Finally, 
 $\vert 0; \varphi_\perp, w_\parallel>$ is the oscillator
ground state, also   characterized  by the  transverse
position $\varphi_\perp$ of the D1-brane,   and by the 
  Wilson line $w_\parallel$ on its worldvolume. To simplify
notation, we will suppress
the dependence on these zero modes  in what follows. 
Neither
the normalization ${\cal N}$ nor the
  zero modes will, in any case, contribute to
the Casimir energy that interests us here.
The reader can  verify easily that 
\begin{equation}
(a_{n}^i -   S_{ij}\; {\tilde a}^j_{-n})\;   \vert \vartheta \gg \ =0 \ .  
\end{equation}
These are precisely  the gluing conditions  \eqref{bcS},
 after `folding' the plane along the defect line.

  In order to calculate the Casimir energy, 
we first Wick rotate  the coordinate $t$ and compactify it on
a circle of circumference $2T$. We also periodically identify   $x \equiv
 x+2L$.
This differs from   the Dirichlet conditions used in  Appendix A,
but the difference 
will go away  in the limit of infinite   $L$ .  
The vacuum  energy 
for the configuration of figure 4 can 
be written as follows in the closed-string channel:
\begin{equation}
{\cal E} = {\underset {T\to\infty} 
{\rm lim}}  -{1\over 2T}\; {\rm log}\; \ll \vartheta \vert \;
e^{-H^1\; 2\pi (L-d)/T }\; e^{-    H^2\;2\pi d/T} \vert \vartheta \gg 
\ ,
\label{jj12}
\end{equation}
with  $H^1$ and $H^2$ the free-field Hamiltonians 
of  $\phi^1$ and $\hat \phi^2$.
The limit    $L\to\infty$  projects
onto  the ground state of $\phi^1$, so that
only the $\hat\phi^2$ oscillators should be kept in the expression
\eqref{bstate} for  the boundary state.
The above matrix element thus becomes
$$
< 0\vert\; 
\prod_{n=1}^\infty {\rm exp}\left(
{1\over n}  {a^{2}_{n}} {{\tilde a^{2}}_n}
\; {\rm cos}\; 2\vartheta\right)\;
e^{- H^2\;2\pi d/T}\;
\prod_{n=1}^\infty {\rm exp}\left(
{1\over n} a^2_{-n} {\tilde a^2}_{-n}\; {\rm cos}\; 2\vartheta  \right)\;
  \vert 0>\;  =  
$$
\vskip -.2cm
\begin{equation}
=\;  {\cal N}^2 \prod_{n=1}^\infty \left(1- {\rm cos}^22\vartheta\;
  e^{-n\; 4\pi d/T}\right)^{-1}\ .
\label{j12}
\end{equation}
Taking the logarithm  converts the product
into a sum, which in the limit $T\to\infty$ reduces to a 
a continuous  integral, 
\begin{equation}
{\cal E} \; = \; 
  {1\over 8\pi d}\int_0^{1}  {dz\over z}\;  {\rm
    log}(1- {\rm cos}^22\vartheta\; z)\ .
\end{equation}
Using, finally,  the integral
 representation of the dilogarithm function \cite{lewin}, 
\begin{equation}
 \int_O^{w}  {dz\over z}\;  {\rm log}(1- z)\;
=\; - {\rm Li}_2 (w)\ , 
\end{equation}
we  recover  precisely the result \eqref{dilog} of  appendix A. 
The dilogarithm function has
 appeared before in the CFT literature \cite{NRT},
but the present  context is, in our opinion, particularly simple. 
The expression  \eqref{j12} for the matrix element
has also 
appeared in the literature before,
under the name `quantum dilogarithm' \cite{Fa,Fa1}.

We can also evaluate \eqref{jj12} for 
$L$ finite.  If we denote $q_1 = \exp (-{2\pi d/ T})$
and $q_2 = \exp ( -2 \pi (L-d) /T)$,
  then the relevant matrix element reads~:
\begin{equation}
{\cal N}^2 \prod_{n=1}^{\infty} 
\left[ 1-   (q_1^{2n} + q_2^{2n})\;{\rm cos}^2  2\vartheta 
 -2 q_1^n q_2^n\; \sin^2 2 \vartheta + q_1^{2n} q_2^{2n} \right]^{-1}\  .
\end{equation}
Sending  $q_2 \rightarrow 0$ gives back the expression   \eqref{j12}
as expected. When  $d={L/ 2}$, on the other hand, the matrix element 
reduces  to ${\cal N}^2 \prod_{n=1}^{\infty} (1-q^{2n})^{-1}$, where
$q=q_1=q_2$. The Casimir energy is independent of
$\vartheta$ in this special case. This is  consistent  with the fact that
the mass subtraction  for a closed string (corresponding to $\vartheta=\pi/4$)
is twice the subtraction for an open string (which corresponds to
$\vartheta = 0$  or $\pi/2$).

 We conclude this section with a brief discussion of  other
gluing conditions, and in particular those corresponding to the matrices 
$M^\prime(\vartheta)$. 
Let $\;^*\phi^1$ be  the field T-dual to $\phi^1$,
which obeys  $\partial_t^{\ *}\phi^1 = \partial_x\phi^1$ and 
$\partial_x^{\ *}\phi^1 = \partial_t\phi^1$. 
It follows from the relation
\begin{equation}
M^\prime(\vartheta)\;  =\;  \left(
\begin{array}{lll}
0 & \quad & 1\cr
1 & \quad & 0 \cr
\end{array}
\right)
   M(\vartheta)\ ,
\end{equation}
that the $M^\prime$ gluing condition describes 
a D1-brane in the direction  $\vartheta$ 
on the  $(\;^*\phi^1, \hat\phi^2)$ plane. The  T-duality 
that takes us   back to the $(\phi^1, \hat\phi^2)$ plane,  transforms
this  D1-brane  into a D2-brane with a  non-vanishing worldvolume
magnetic flux  \cite{Brev}. In the simplest case of 
a compact scalar  and a diagonal D1-brane,  
as in figure 5, the T-dual configuration is characterized by  one unit of 
magnetic flux. 
We should stress,  however, that the relation 
\eqref{theta} between the angle $\vartheta$ and the radii is consistent,  but
by no means  unique. It was derived from the hypothesis that the field
$\tilde\phi$ of section 2.2 should be  continuous across the wall. 
A  more general consistent  hypothesis is that 
$ \tilde\phi(-0) = n  \tilde\phi(+0)$,  leading  to
 the relation 
\begin{equation}
{\rm tan}\;\vartheta \; = \;\lambda\; =\;  { r_2\over n  r_1}\ \; .
\label{theta1}
\end{equation}
This corresponds  (after folding)  to a 
 D1-brane that winds $n$ times  around 
dimension 1, but  only a single time  
 around   dimension 2. The T-dual configuration
is a D2-brane carrying   $n$ units of magnetic flux. 
As  will 
become  in fact clear  in the following section, 
 {\it any} consistent D-brane
configuration on the two-torus can be 
`unfolded' to a conformally-invariant interface of the one-scalar theory.

\vskip 0.3cm
 

\section{Supersymmetry and generalizations}

  The analysis of the previous section  can be extended in several 
directions.
One may consider  abstract gluings  of conformal theories, mutliple
interfaces or junctions,  Calabi--Yau sigma models,  or orbifold theories. 
Another important question  concerns 
 the supersymmetry  properties of
the walls. In this section we will elaborate
 on some of these various issues.


\subsection{Fermions and supersymmetry}

The $N=(1,1)$  supersymmetric extension of the free-scalar  model
 has  a pair
$(\psi_+, \psi_-)$ of  Weyl-Majorana fermions, which are 
the superpartners  of the field $\phi$. 
Conformal invariance 
requires continuity of $(T_{++} - T_{--})$ for  the fermions. 
Supersymmetry, on the other hand, further requires that 
\begin{equation} \label{j0}
 ( G_+ +  \eta  G_-) \Bigl\vert_{-0}
 = \pm \;  ( G_+ \pm  \eta  G_-) \Bigl\vert_{+0} \ ,  
\end{equation}
where $G_\pm $ are  the left and right supercurrents, and $\eta=\pm 1$.
For a single wall, the
three sign ambiguities in this condition can be absorbed
in redefinitions of the  fermion fields. 
The signs involving  only the fields on the same side of the
wall  are basically  irrelevant (except possibly if $x$ is
compactified) and we will henceforth take them  to
be positive.  
The third sign,  $\eta$, on the other hand, involves fields on
both  sides of the wall, and will therefore be 
important when two or more
interfaces are present. As we will see, $\eta$  distinguishes
an interface from an anti-interface.

In order to make the supersymmetry manifest, we will show how
these boundary conditions arise directly in superspace. Consider
 the general $N=(1,1)$ supersymmetric sigma model with action
\begin{equation} \label{j1}
I  = \int dx\; dt\;  d^2 \theta\;  \left[ G_{IJ}(\Phi) +
 B_{IJ}(\Phi)\right] \;  D_+ \Phi^{I} D_- \Phi^{J}\ , 
\end{equation}
where
\begin{equation}
  D_\pm  = \frac{\partial}{\partial \theta^\pm  } +
 \theta^\pm  \left(\frac{\partial}{\partial t}  \pm 
 \frac{\partial}{\partial x} \right) .
\end{equation}
If this sigma model is the CFT on the left of the domain
wall, we need to find the variation of the action,  and match
it to the corresponding variation on  the right side of the wall. 
We assume here, as we did until now, that the domain
wall does not support any independent degrees of freedom.

The variation of the action (\ref{j1}) yields the following
boundary term: 
\begin{eqnarray}
\delta I  & = & 
- \int dt\;  \left[ \;\frac{1}{2}
 \Sigma_{J}\;  \delta(D_+\Phi^{J} + D_-\Phi^{J})  +
\frac{1}{2} (D_+ + D_-) \Phi^{J} \delta \Sigma_{J} \right. 
\nonumber \\
& & \qquad \left.
 -\; \delta \Phi^{J} (D_+ + D_-)\Sigma_{J}
 \Biggr]_{x=-0,\; \theta^\pm =0}  \right. 
\ \  , \label{j2}
\end{eqnarray}
where
\begin{equation} \label{j3}
\Sigma_{J} = [G_{JK} + B_{JK}]\;  D_+ \Phi^{K}
 - [G_{JK} + B_{JK}]\;  D_- \Phi^{K}\ .
\end{equation}
In deriving equation
 \eqref{j2}
 we used the equation of motion
for the auxiliary field, which is the top component of the
superfield $\Phi$. In addition, we have dropped 
the variation of a pure boundary term,  $-\delta I_b$ , with
\begin{equation} 
I_b = \frac{1}{2} \int dt\;   B_{IJ}(\Phi) \left(
D_+ \Phi^{I} D_+ \Phi^{J} +
D_- \Phi^{I} D_- \Phi^{J} \right).
\end{equation}
This  is of course absent for $B_{IJ} =0$, and in particular
if there is only one superfield. 
More generally, we should have included this boundary term in the
action (\ref{j1}) in order to
arrive at the above variation.

The form of the variation (\ref{j2}) suggests  that we introduce
 two `half' superfields \cite{Hori} as follows: \footnote{
See also \cite{ulf} for a recent detailed
 analysis of  supersymmetry-preserving 
boundary conditions in general $N=(1,1)$  sigma-models.}
\begin{equation}
\tilde{\Phi}^{J}(x,t,\theta) =
 \Phi^{J}\Bigl\vert_{\theta^+=\theta^-=\theta}\qquad
{\rm and}  \qquad
\tilde{\Sigma}_{J}(x,t,\theta) =
\Sigma_{J}\Bigl\vert_{\theta^+=\theta^-=\theta}\  .
\label{half}
\end{equation}
For example, in flat space
 and for zero  $B_{IJ}$ we find: 
\begin{equation}
\tilde{\Phi}^{J} = \phi^{J} + \theta\;(\psi_+^{J} + 
\psi_-^{J}) \qquad {\rm and}  \qquad
\tilde{\Sigma}_{J} = (\psi_+^{J} - 
\psi_-^{J}) + 2 \theta\;  \partial_{x} \phi^{J}\  .
\end{equation}
One can now verify easily that,  if these half superfields are 
 continuous across the
wall, then the variations \eqref{j2} of the left and right CFTs
will precisely cancel out  each other.
 In addition, a manifest
$N=1$ supersymmetry will be preserved, since everything can be 
expressed in terms of half
 superfields. The superderivate in
this half superspace is defined as:
\begin{equation}
D \equiv  D_+ + D_- = \partial_{\theta} + 
2 \theta \partial_{t}\ , 
\end{equation}
 and  since it
does not contain a derivative of $x$, it acts indeed  along
the interface. Another way of arriving at the above  conclusion, is
by constructing  the superfield combination 
\begin{equation}
\Theta\; \equiv \; 
 \frac{1}{8}\left[  D^2 \tilde{\Phi}^{J} \; \tilde{\Sigma}_{J}\; + \;
  D \tilde{\Phi}^{J}\; D\tilde{\Sigma}_{J}\right] \; 
= G_++G_-\;  +\; \theta\; (T_{++}-T_{--})\ 
.\label{superem}
\end{equation}
From this one sees immediately
 that continuity of the  half superfields \eqref{half}
 across the wall
  implies, indeed,  the boundary conditions
given in (\ref{j0}), with $\eta$ (and all the other signs) 
chosen to be positive. 

  The choice $\eta=-1$ corresponds to another set of half-superfields,
which are obtained by setting $\theta^+ = -\theta^- =\theta$.
The combination \eqref{superem}, with $D\equiv D_+-D_-$,
has now $G_+-G_-$ as  its lowest component (and the same
upper component as above). Continuity of
this new set of half-superfields  respects, therefore, the $\eta=-1$
superconformal-invariance conditions. Two interfaces with opposite
values of $\eta$   break completely all the supersymmetry.

If there are  more than one superfield, the vanishing of \eqref{j2}
is guaranteed by the more general  boundary conditions
\begin{equation}
\left(
\begin{array}{l}
\tilde{\Sigma}_{J} \cr
D\tilde{\Phi}^{J} \cr
\end{array}
\right)_{x=-0}
=\;\;
M \;
\left(
\begin{array}{l}
\tilde{\Sigma}_{J} \cr
D\tilde{\Phi}^{J} \cr
\end{array}
\right)_{x=+0}
 \ \ ,
\label{bcS}
\end{equation}
with the constant matrix $M \in O(d,d)$. 
Of course, our discussion here is entirely
 classical, and superconformal
symmetry could   be broken by quantum corrections. 
Furthermore, one needs to check compatibility of the above conditions  
with the global structure of the target space of the sigma
model. Thus, in general,   only a limited subset of $O(d,d)$ gluings will 
be allowed.


   The calculation of the Casimir energy of the previous section can
be extended easily to the superconformal case. 
The gluing  conditions for the fermionic fields  that supersymmetrize
equation  \eqref{bcS} are:
\begin{equation}
\left(
\begin{array}{l}
\psi_-^1\cr
\psi_+^2\cr
\end{array}
\right)
=\;\;
S (\eta)\; 
\left(
\begin{array}{l}
\psi_+^1\cr
\psi_-^2\cr
\end{array}
\right) \ \ , 
\label{bcFS}
\end{equation}
where
\begin{equation}
S (\eta) =   \; \left(
\begin{array}{lll}
 -\eta\;  {\rm cos}\;2\vartheta & \quad &  \ \;{\rm sin}\;2\vartheta \cr
\ \ \;\; {\rm sin}\;2\vartheta & \quad & \eta\; {\rm cos}\;2\vartheta \cr
\end{array} 
\right) \  ,  
\label{SOF11}
\end{equation}
with a similar expression for $S^\prime$. The factors of $\eta$
in the gluing matrix follow from the fact that 
changing  $\eta$ is the same
as  flipping the sign of the $\psi_-^j$. 
The fermionic part of the boundary state
 that imposes these gluing conditions is
\begin{equation}
\vert \vartheta , \eta  \gg _F \;  =\  {\cal N}^\prime\;
\prod_{r  > 0} {\rm exp}\left(i\;
\psi^i_{-r} {\tilde \psi^j}_{-r} \;  S_{ij}(\eta)  \right)\;  \vert 0 > \
.
\label{bFstate}  
\end{equation}
The factor of $i$ in the exponent arises in  going from
the open to the closed-string channel \cite{ga}, and 
${\cal N}^\prime$ is  a (irrelevant for us) normalization. 
The frequencies $r$ can be either integer or half-integer, depending on
whether we are in the  Ramond or Neveu-Schwarz sector of the closed-string.

Proceeding as in section 2.3 we obtain  the following expression
for the Casimir energy:
\begin{equation}
{\cal E} = {\underset {T\to\infty} 
{\rm lim}}  -{1\over 2T}\;    {\rm log}\; 
 { \prod_r\; (1 -\eta_L\eta_R \;  {\rm cos}^22\theta\;
  e^{- r \; 4\pi d/T}) \over 
\prod_n\;  (1- {\rm cos}^22\theta\;
  e^{-n\; 4\pi d/T}) 
} \  .
\end{equation}
Since $T\to\infty$,  the result does not
 depend on the choice of integer or half-integer $r$.
What does make a difference is whether the left and right interfaces
are of the same  or of opposite type:  $\eta_L\eta_R=+1$ or $-1$.
In the first case supersymmetry is preserved and the Casimir energy is zero.
In the second case one finds that
\begin{equation}
{\cal E} \; = \;  -{1\over 8\pi d} \; \left[ Li_2({\cal R}^2)
 -  Li_2(- {\cal R}^2) \right]\ 
\; = \; -{1\over 8\pi d} \; [ 2\;  Li_2({\cal R}^2)
 -  {1\over 2} Li_2( {\cal R}^4) ]
\ .  
\end{equation}
The last  equality  follows from a
 standard  dilogarithmic identity.
Writing ${\cal E}$ in this form
 shows that  in the case of total reflection (${\cal R} = \pm 1$)
the result is $3/2$ times the bosonic contribution. This  is indeed the
 vacuum energy of  a superfield  
with conventional Neveu-Schwarz boundary conditions.
For weak reflection, on the other hand, the bosonic  and fermionic
contributions to the Casimir  energy are equal.


Let us  finally discuss  $N=(2,2)$ sigma
models with a target space that is a K{\"a}hler manifold. In this case, 
the sigma-model action takes  the form
\begin{equation}
I  = \int dx\; dt\; d^4 \theta \; K(\Phi^i,\bar{\Phi}^{\bar{i}})\ ,
\end{equation}
where $\Phi,\bar{\Phi}$ are (anti)chiral superfields that
satisfy $\bar{D}_\pm \Phi^i = D_\pm  \bar{\Phi}^{\bar{i}}=0$.
The superderivatives are: 
\begin{equation}
D_\pm  = \frac{\partial}{\partial \theta^\pm  } 
 + 2 \bar{\theta}^\pm  \partial_\pm \  , \qquad
\bar{D}_\pm  = \frac{\partial}{\partial \bar{\theta}^\pm  } 
 + 2 {\theta}^\pm  \partial_\pm  \  .
\end{equation}
Repeating
 our previous  analysis, we
 find  that the variation of the action can be
written again in terms of half superfields. The relevant half
superfields now are: 
\begin{eqnarray}
 {\varphi}^{i}(x,t,\theta ,\bar{\theta})\; & = &\; 
 \Phi^i \Bigl\vert_{\theta^+=\theta^-=\theta ,\; 
 \bar{\theta}^+=\bar{\theta}^-= \bar{\theta}} \ ,  \nonumber \\
 {\bar{\varphi}}^{\bar{i}}(x,t,\theta ,\bar{\theta})\; & = &\; 
 \bar{\Phi}^{\bar{i}} \Bigl\vert_{\theta^+=\theta^-=\theta ,\; 
 \bar{\theta}^+=\bar{\theta}^-= \bar{\theta}}\ ,   \nonumber \\
\Lambda_i(x,t,\theta ,\bar{\theta})\; & = &\; 
\partial_i \partial_{\bar{j}} K (\bar{D}_+ - \bar{D}_-)
 \bar{\Phi}^{\bar{j}}\Bigl\vert_{\theta^+=\theta^-=\theta ,\;
 \bar{\theta}^+=\bar{\theta}^-=\bar{\theta}}\ ,  \nonumber \\
\bar{\Lambda}_{\bar{i}}(x,t,\theta ,\bar{\theta})\; & = &\; 
\partial_i \partial_{\bar{j}} K ({D}_+ - {D}_-)
 {\Phi}^{{i}}\Bigl\vert_{\theta^+=\theta^-=\theta ,\;
 \bar{\theta}^+=\bar{\theta}^-=\bar{\theta}}\   . \label{j8}
\end{eqnarray}
The half-superspace  coordinates are $\theta $ and
$\bar{\theta}$, with derivatives $D = (\partial_{\theta } + 
2 \bar{\theta} \partial_{t})$ and 
$\bar{D}= (\partial_{\bar\theta} + 
2 \theta  \partial_{t})$.
By requiring the above half superfields to be continuous accross
the domain wall, we automatically preserve
 one $N=2$ algebra.
The generators of this  algebra are the components of the half superfield
\begin{equation}
\label{half2} 
 (D  {\varphi}^i) \Lambda_i + \bar{\Lambda}_{\bar{i}} \bar{D}
{\bar{\varphi}}^{\bar{i}} \ .
\end{equation}
These  are clearly continuous across the wall,   once the fields
in (\ref{j8}) are themselves  continuous. Note that the lowest component of
\eqref{half2}  is the difference of the left- 
 and right-moving  $U(1)$  currents.

More generally,  if the target
space is $d$-complex-dimensional, there is an $O(d,d,{\bf C})$ 
family of candidate boundary conditions.
 The subgroup $GL(d,{\bf C})
\subset O(d,d,{\bf C})$ has a  simple interpretation in terms
of holomorphic branes in ${\cal M}_L\times {\cal M}_R$, where
${\cal M}_{L,R}$ are the two target manifolds
 on either side of the interface.
Indeed, let  $v^i$ be complex coordinates for ${\cal M}_{L}$
and  $w^i$  complex coordinates for ${\cal M}_{R}$.
Then $v^i=A^i_j w^j$ defines, in a local patch, a
holomorphic $d$-complex dimensional brane.
When this  brane can be defined globally (we will discuss such
an example in the following  subsection) then 
it gives rise to a $N=2$ superconformal interface.
Since holomorphic branes  are BPS, 
we expect them to survive in the
quantum theory, at least in the  large-volume limit.


\subsection{Generalizations }
The folding trick allows us to discuss
conformal-field-theory gluings more  abstractly.
Start with the tensor product
of two  conformal theories, CFT1$\otimes$CFT2,
defined on the Euclidean half plane,  $Imz\geq 0$. The two theories
need not be identical, nor even have equal  central charges.
Conformal  boundary conditions are   described by a 
boundary state $\vert {\cal B}\gg $,
which satisfies 
\begin{equation}
\label{conti}
\left( L^{(1)}_n +  L^{(2)}_n -  \tilde{L}^{(1)}_{-n}
- \tilde{L}^{(2)}_{-n} \right)\vert {\cal B}\gg \ =\  0 \ .
\end{equation}
Here $L^{(1)}_n$ and $\tilde L^{(1)}_n$
are the left-moving, respectively  right-moving Virasoro generators of CFT1,
in the closed-string channel,  
and similarly for CFT2. 
If we `unfold' CFT2
 unto the lower half-plane,  $Imz\leq 0$, 
the roles  of its
holomorphic and antiholomorphic fields  are interchanged. 
Condition \eqref{conti}
 then precisely 
ensures the continuity of $T_{zz} - T_{\bar z \bar z}$
on  the real axis. 
In this way,  any conformal boundary state in the tensor-product  theory 
can be unfolded into a conformal interface, and vice-versa.

   A trivial situation arises whenever the boundary state can be factorized,
\begin{equation}
\label{facto}
\vert {\cal B}\gg _{\rm reflect} \; =\;
 \vert {\cal B}_1\gg \; \otimes\; \vert {\cal B}_2\gg \ .
\end{equation}
In this case $L_n - \tilde{L}_{-n}$ vanish for each theory separately, 
so that $T_{xt}$  is zero at the interface. 
There can,  therefore,  be no transfer of energy across the wall, and the
two conformal field theories
 are 
 decoupled.\footnote{Any linear combination of states
of type \eqref{facto} will,  likewise,  give   perfect reflection.
By an abuse of language, we  keep
refering to such states as `factorizable', since 
the two conformal theories dont talk,  except possibly
via  correlated boundary conditions.}
At the opposite end of the spectrum are the perfectly-transmitting
defects, for which
\begin{equation}
\label{conti2}
\left( L^{(1)}_n - \tilde{L}^{(2)}_{-n}
 \right)\vert {\cal B}\gg _{\rm transmit} \ =\ 
 \left( L^{(2)}_n - \tilde{L}^{(1)}_{-n}
 \right)\vert {\cal B}\gg _{\rm transmit} \ =\ 0 \ .
\end{equation}
Such states obviously exist when the two CFTs are identical, but not only.
For instance, for the scalar field of section 2  one may consider a D1 brane
at $45^o$,  even if the radii on the two sides of the
interface are not the same. Generic  permeable defects  are those
for which  the boundary  state $\vert {\cal B}\gg $  is of neither of the above
two special types.

  As well-known, the Virasoro gluing equations \eqref{conti} must be
supplemented, in general,  by global consistency conditions (for
reviews  and references see \cite{SS,SF,Zu}). For instance,  the
annulus diagram must be a partition function with integer
multiplicities in the open channel \cite{Ca}. Such conditions will be
obeyed automatically for defects with  a local action principle,
like those we have considered up to  now. 
From a more algebraic point of view, it will be  sufficient to ensure that
the  state $\vert {\cal B}\gg $ in  the tensor-product  theory is consistent. 
The consistency of the bulk and boundary operator algebra can be,
indeed, verified before the procedure
 of `unfolding'.  Notice, in particular, that
the  operators living on the interface are the local boundary operators 
allowed by the state $\vert {\cal B}\gg $. 

\begin{figure}[ht]
 \hskip 3.0  cm
       \hbox{\epsfxsize=80mm%
       \hfill~
       \epsfbox{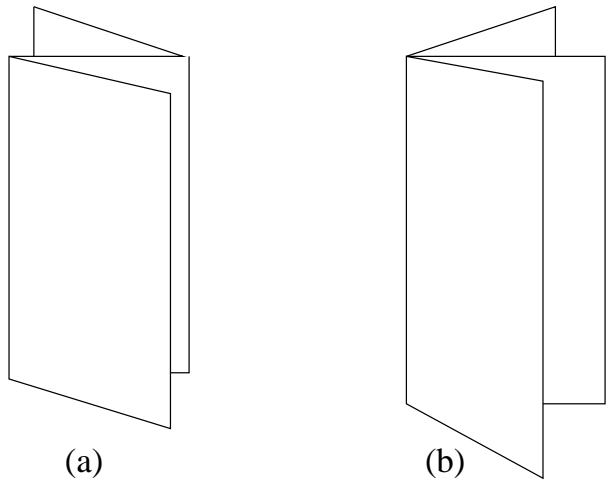}
       \hfill~}
       \caption{The folding of (a) two neighbouring interfaces, and
(b) a triple junction of conformal theories. }
\end{figure}

  These considerations can be generalized easily to any number of
adjacent parallel  defects.  
One  must  fold along the interfaces  repeatedly, as illustrated 
in  figure 6a, so as to make  an annulus with many sheets. 
The boundary
conditions at the folds are  
boundary states 
of  the  product theory  
${\rm CFT1} \otimes \overline{\rm {CFT2}} \otimes \cdots {\rm CFT}k$,
where CFT$m$  is the theory  on the $m$th  sheet
and for even $m$ the left-
 and the right-movers must be exchanged (this is indicated here
by a bar).  Note that one can introduce 
extra  folds with  purely-transmitting boundary  conditions. 
One can also consider multiple junctions of CFTs, as
illustrated in figure 6b (for an earlier  study of  field
theories  on string junctions see \cite{CT}).  
Extending the calculations of the previous section in such  contexts
is a straightforward exercise that we do not pursue.

  The construction of  permeable interfaces of  strongly-interacting CFTs
is  a very  interesting question,  to which we hope to return
in  future work.
Here,  we  want to conclude our discussion with a few more
simple examples of  domain walls. First, let us consider
the case of several free scalar fields, $n_1$ on the left and $n_2$ on
the right of the interface.
 The boundary states are 
(combinations) of planar branes in $n_1+n_2$ dimensions, 
 which are generically at
angles and can carry a magnetic flux. 
 If the scalar fields 
have canonically normalized 
stress tensors,   
the gluing conditions will be of the same form as (2.7), 
with   $S$  an orthogonal  matrix that we write in terms of
$n_i\times n_j$ blocks:
\begin{equation}
S  =   \; \left(
\begin{array}{cc} S_{11} & S_{12} \\ S_{21} & S_{22} \end{array}
\right)\ .  
\end{equation}
Repeating the  Casimir-energy  calculation
of section 2 gives:
\begin{equation}
{\cal E} = - \frac{1}{8\pi d}\; {\rm Tr}\; 
[ {\rm Li}_2\;(S_{22}^{\ 2})].
\end{equation}
Notice that 
the pressure on the walls  only depends, as should be expected, 
on the reflection amplitudes  of the conformal theory CFT2
that lives in the space 
in between these walls.

\begin{figure}[ht]
 \hskip 3.0  cm
       \hbox{\epsfxsize=105mm%
       \hfill~
       \epsfbox{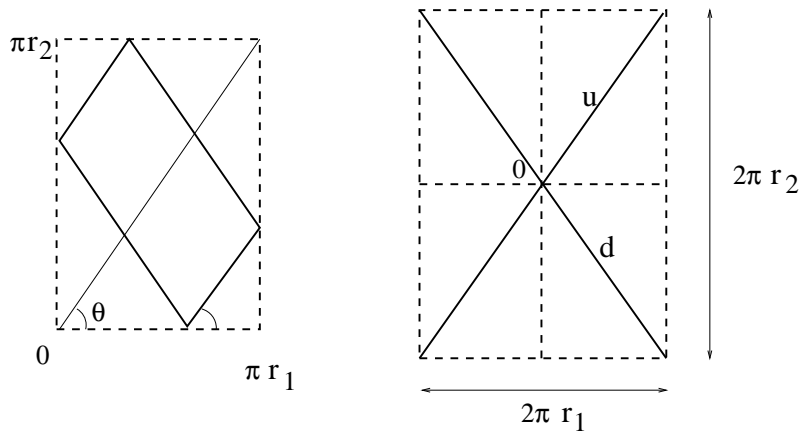}
       \hfill~}
       \caption{
A regular  D1-brane  of the S1/Z2$\times $S1/Z2 orbifold theory that
winds once around each of the covering circles (left). When forced to go
through the origin, this D-brane has a single,
 rather than  three, images under
reflections (right). In this case it can 
 decompose (assuming also vanishing Wilson line)   into
two,~more elementary,  fractional branes. }
\end{figure}

For  a less trivial example, let us discuss orbifolds. 
Consider  the case where on either
side of the interface 
lives a $c=1$ orbifold theory, so that the tensor product
CFT  has target space  S1/Z2$\times $S1/Z2. 
A  D1 brane  winding once around each of the
two covering circles has the generic 
form shown in  figure 7~.  It is an inscribed parallelogram,
with sides  parallel to the two diagonals of the target space.  
There is, furthermore, 
a four-fold Chan--Paton multiplicity, corresponding to
the four images of the D-brane in the covering torus. 
Marginal deformations change
the shape of the parallelogram, while keeping  its  four angles fixed,
and also turn on a Wilson line.
At  special point(s) of this  moduli space, where 
 the  parallelogram collapses  along  a diagonal 
of target space,  as in figure 7, 
 the    D1-brane 
decomposes   into two,  more elementary,
 fractional D-branes \cite{frac,frac1,frac2}. 
These are the basic branes of the tensor-product theory  
which, 
in the limit of equal  radii ($r_1=r_2$),  unfold into
perfectly-transmitting interfaces. 
 

   We can  extend
the above discussion 
to   $N=(2,2)$ supersymmetric sigma models on  orbifold spaces,
like T4/Z2 or T6/Z3. Consider the latter example which is a
(singular) Calabi-Yau surface with a unique complex structure
and 36 K{\"a}hler moduli. Varying the nine untwisted moduli
separately, for the two sigma models of the tensor product,
will lead to diagonal branes that describe permeable interfaces.
Vaying the 27 twisted moduli will blow up some of the orbifold fixed
points. Since the complex structure is here unique, we 
expect the middle-dimensional holomorphic branes 
described in section 3.1. to survive


\section{The NS5/F1  system and holography}

We will now  apply  the ideas of CFT domain
walls to branes in   AdS3. Of special  interest to us
are the static  one-branes extending  all the
way to spatial infinity \cite{BP}.
Since these are codimension-one in the bulk, 
they  separate two different supergravity vacua, 
distinguished by their charges.  Correspondingly on the boundary we
find $0+1$ dimensional domain walls  separating  two different CFTs,
that a priori can have different central charges. Since  the stable
one-branes are supersymmetric and 
have AdS2 geometry \cite{BP},  the corresponding domain walls
should  be superconformal. 

One way of trying to test  this correspondence is by comparing 
the Casimir energy of the  walls, both from  the supergravity and 
from the CFT viewpoints. This is
the two-dimensional analogue of the Wilson loop calculation \cite{qq1,qq2} 
in  four dimensions. It is also one  version of the more general
Karch-Randall setup \cite{KR} in which 
two $n$-dimensional CFT's are glued together with a
$(n-1)$-dimensional CFT.

\subsection{String theory setup}

Our starting point is the type IIB string compactification on 
a four-manifold $M^4$,  which   is either a
four-torus or a K3  surface.
The resulting
six-dimensional theory  contains a variety of  strings. Besides
the fundamental string and D-string of the
 uncompactified IIB theory,
there are also D3 branes 
wrapping the various two cycles of $M^4$, as well as  
 D5  and NS5 branes  wrapping the entire manifold.
The  strings are labeled by a charge vector $\vec q$ in  the lattice
$\Gamma^{5,5+n}$,  where $n=0$ for $M^4=T^4$ and $n=16$ for $M^4=K3$.
Furthermore, 
 there is a $O(5,5+n,{\bf Z})$ duality group which permutes
the  different charges,  keeping the  invariant length $\vec q^{\; 2}$ fixed.
 The moduli space of this string compactification is
\begin{equation}
O(5,5+n,{\bf Z})\backslash O(5,5+n)/O(5) \times O(5+n)\ .
\end{equation}

We can distinguish two classes of BPS strings. First, those  with a
(primitive)  charge vector of zero length,   $\vec
q^{\; 2}=0$,  which  lie  in the $U$-duality orbit of the
 fundamental string. 
Such objects are weakly coupled
in some  corner of the moduli space, 
and can be chosen as the fundamental quanta  in
a perturbative expansion. 
Secondly, there are strings with 
 $\vec q^{\; 2}$  positive.\footnote{For $M^4=T^4$
the negative  $\vec q^{\; 2}$ strings are also supersymmetric.}
These  can  be always mapped,  by a $U$-duality transformation,  to a bound
state of  $Q_1$ fundamental strings and $Q_5$ NS
fivebranes, where 
\begin{equation}
 \vec q^{\; 2}= 2 Q_1 Q_5 
\end{equation}
If the charge vector $\vec q$ is furthermore primitive,  $Q_1$ and $Q_5$ are
relatively prime and we have a well-defined bound state.
We want to study the near-horizon decoupling limit for  such a configuration.
The  relevant geometry is AdS3$\times $S3$\times M^4$, and the dual
 supersymmetric CFT
has  total central charge $6N = 6\;Q_1Q_5$ .

Picking  a particular charge vector $\vec q$, 
reduces the duality group and moduli space. 
The remaining $U$-dualities,
 that are realized as $T$-dualities in the CFT, are given by the `little
group'  $O(4,5+n,{\bf Z})$ that preserves the charge vector $\vec q$. By the
attractor mechanism \cite{Fe1,Fe2}
some of the scalar fields that parametrize the
moduli space take specific fixed values.  More explicitly,  if we use the
Narain decomposition $\vec q=\vec q_L + \vec q_R$  with $\vec q^2= \vec
q_L^2- \vec q_R^2$, the attractor equation gives $\vec q_R=0$. The
moduli space of the supergravity solution
is then reduced to the  homogeneous space
\begin{equation}
\label{fundd}
O(4,5+n,{\bf Z})\backslash O(4,5+n)/O(4) \times O(5 +n)\ .
\end{equation}
Note that $\vert\vec q_L\vert$ is the tension of the background string.
Note  also that the full parameter  space of the dual (spacetime)  CFT 
includes many copies of the  `fundamental domain' \eqref{fundd},   and has an 
intricate  global structure  \cite{LM}.

This six-dimensional  theory contains various string junctions
where a string with charge $\vec q_1$ absorbs a string with charge
$\vec q_2$ to form a string with charge $\vec q_1 + \vec q_2$.
The superconformal walls are holographic duals of such junctions.
We will choose a duality frame
 where the background $\vec q_1$ is built out of only
fundamental strings and NS fivebranes. Its near-horizon geometry
carries, therefore,  Neveu--Schwarz  fluxes only. 
The full type-IIB
string theory can  be described  in this case by a 
Wess--Zumino--Witten  model on the group manifold  $SL(2,{\bf
R})\times SU(2)$, together with a sigma model with $M^4$ target space.
The $Q_1$-dependence appears through the six-dimensional
string coupling, which is fixed by the attractor mechanism to be
\begin{equation}
{1\over g_6^2} = {Q_1\over Q_5}\ .
\end{equation}
For a reliable supergravity approximation  one needs
therefore   $Q_1\gg Q_5\gg 1$. 

  Let us consider now a  second string with charge vector
$\vec q_2$ , stretching   between two points, 
$x=0$ and $x=2R$,   on the AdS3 boundary  as in figure 8.
In the dual holographic field theory  the string endpoints are  a wall and an
antiwall, separating two different CFTs. 
With the use of $T$-dualities
we can map this  second string to a configuration that does not contain
D3-branes. Although we will mostly work with $(p,q)$ strings below, the most
general configuration can also involve D5-  and NS5-branes. The
$U$-dualities that preserve the vector $\vec q_1$ are,
  in general,  insufficient to
always map $\vec q_2$ to only fundamental strings and D-strings.

\vskip 0.2cm


\begin{figure}[ht]
       \hbox{\epsfxsize=140mm%
       \hfill~
       \epsfbox{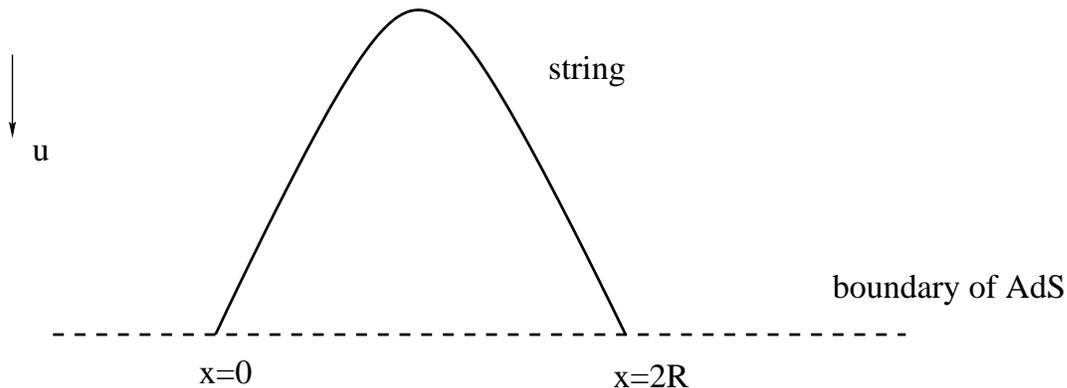}
       \hfill~}
       \caption{Stretched string between a wall and an  antiwall.}
\end{figure}

A $(p,q)$ string like the one of figure 8  will only equilibrate
if one applies a force to keep   its two endpoints from
collapsing.
From the  holographic point  of view, this force is  the
Casimir attraction of the walls. We will now compute it in the
supergravity approximation.   
In order to do  a reliable  calculation we 
assume that the tension $T_{(p,q)}$ of the probe  string
 is much smaller than the tension $T(\vec q_1)$ of the background
 string, so that backreaction can
be consistently neglected. 

  The calculation is similar in spirit to the Wilson-loop calculation
in the supergravity limit of $N=4$
 super-Yang-Mills \cite{qq1,qq2}. The string coupling 
to the background $B$-field  
introduces, however, a new parameter at the technical level.

\subsection{Supergravity  calculation}

The metric and B-field backgrounds of the $SL(2,{\bf R})$ WZW model in
Poincare coordinates are
\begin{equation}
ds^2 = L^2 \left[ \frac{ du^2}{ u^2} +   u^2  (dx^2 - dt^2)\right]
\ \ \ \ \ {\rm and}\ \ \ \ \
B = L^2 u^2\;  dx\wedge dt\
\end{equation}
We denote, for short, by  $T$ and  $\rho$ the tension and NS charge density
of the $(p,q)$ string. 
Its  energy,  
as measured by an observer sitting at radial position
$u=1$, takes the form \cite{BP}
\begin{equation}
{\cal E}  = 2L\:  \int_0^R dx\; 
\left[\; T \sqrt{u^4+ u^{\prime\ 2}} - \rho \;  u^2
 \right]\ , 
\label{e}
\end{equation}
with $u^\prime = du/dx$. 
Extremizing leads to the constant of motion
\begin{equation}
  \left[\;
  \frac{T \; u^{4}}{\sqrt{u^4+ u^{\prime\ 2} }}\; -\;
 \rho \; u^2\;  \right]\ 
\equiv \rho C.
\label{c}  
\end{equation}
Setting $C=0$ 
corresponds to free boundary conditions
at the endpoints. The  string   falls,
 in this case,  towards the Poincar{\'e} horizon
and never  comes  back. Its worldsheet  has AdS2 geometry.
More generally,  $C$ and  $R$ are  related implicitely by
\begin{equation}
R = \int_0^R dx = \int_{\infty}^{u_0} \frac{du}{u^\prime}\ \ \ ,
\end{equation}
where  $u_0$ is the minimum value of $u$ , corresponding to $u^\prime =0$. 
Solving \eqref{c} for $u^\prime$ , and making the
change of  variables   
$w\equiv 1/u^2$,  gives 
\begin{equation}
R = \frac{1}{2} \int_0^{w_0} \frac{dw}{\sqrt{w}}\frac{Cw+\rho}{ 
\sqrt{(T+\rho+C  w)(T-\rho-C w)}}\ ,
\end{equation}
with  $w_0= \frac{T-\rho}{C}$.  Performing the integrations  we find
\begin{equation}
\label{Constant}
\sqrt{C}=  \frac{\sqrt{2T}}{R} \left({\bf E}(k) -
 \frac{1}{2}{\bf K}(k)\right)\ ,
\end{equation}
where ${\bf E}$ and ${\bf K}$ are the complete elliptic integrals,
\begin{equation}
{\bf E}(k) = \int_0^\frac{\pi}{2}da\; \sqrt{1-k^2{\rm sin}^2a} \ \ ,\ \ \ \ 
{\rm and } \ \ \ \ 
{\bf K}(k) = \int_0^\frac{\pi}{2} \frac{da}{ \sqrt{1-k^2{\rm sin}^2a} }\ \ .
\end{equation}
The  argument of these integrals is a function of the  tension and the
NS-charge density
of the probe string, 
\begin{equation}
k^2 = \frac{T - \rho}{ 2T}\ \ .
\end{equation}
Equation \eqref{Constant}
expresses  the integration constant  $C$ in terms of  the separation
of the string endpoints.

\vskip 0.3cm
  Let us next evaluate the
 energy. Substituting  $u^\prime$
in equation \eqref{e} and changing again  variables to $w=1/u^2$, 
 leads to the expression:
\begin{equation}
{\cal E} = L \int_{\epsilon^2}^{w_0} \frac{dw}{ w \sqrt{w}}\frac{
T^2 - \rho^2 - C\rho w}{ 
\sqrt{(T+\rho+C  w)(T-\rho-C w)}}\ .
\end{equation}
The integral diverges in the $w\to 0$ limit (near the boundary of AdS)
and has been therefore cutoff at $u= 1/\epsilon$.
  Performing the
integration  gives the following result for the energy:
\begin{equation}
{\cal E} = - L\sqrt{2TC}\; \Bigl[ 2 {\bf E}(k) -
 {\bf K}(k)\Bigr] + {2L\over \epsilon}\; \sqrt{T^2-\rho^2}  \ .
\end{equation}
The divergent second term 
is independent of the distance between the string
endpoints. It could be removed by adding a  boundary term to
 the DBI action, and can be anyway considered
as a renormalization of  the `mass'  of the domain wall. Removing this
divergent term,  and plugging in the expression \eqref{Constant} for the
integration constant, 
leads to the final  expression for the renormalized energy:
\begin{equation}
{\cal E}_{\rm ren} = - \frac{ LT}{ R} \Bigl[ 2 {\bf E}(k) -
{\bf K}(k)\Bigr]^2\ .
\end{equation}
Notice that it  has the correct $1/R$ scaling 
behaviour of  a Casimir energy. This is reassuring, though hardly
surprising. 

  The really interesting story in the above expression is its
non-trivial dependence on $p$ and $q$. This is due to the fact
that the brane has non-trivial coupling to the background flux.
In the standard conventions
in which the ratio of the F-string to the D-string tension is
the string coupling, $g_s$, one finds for the argument of
 the elliptic integrals:
\begin{equation}
\label{ellmod}
2k^2 = 1 - {qg_s\over\sqrt{p^2+ g_s^2 q^2}}\ .
\end{equation}
There are two instructive limits one can consider. First, the limit
$q\to\infty$ (or equivalently $p\to 0$) where the brane
is basically a collection of $q$ pure fundamental strings,
and $k\to 0$.
 In
this limit, the Casimir energy reads
\begin{equation}
{\cal E}_{\rm ren} = - \frac{\pi}{ 8L R}\;  q Q_5 \ ,
\label{q-casimir}
\end{equation} 
where we have used the relation between the background radius and the
number of NS fivebranes,  $L^2= Q_5\alpha^\prime$.
This is the Casimir energy of a CFT with central charge 
$6qQ_5$, confined to  an interval of size $LR$. 
We will explain in the following subsection why this agrees with
the naive sigma-model expectation.

The second interesting limit, that of pure D-strings, is the
natural starting point of a 
 perturbative expansion at  weak string coupling.
From equation \eqref{ellmod} we get:
\begin{equation}
\label{ellmod}
k  = \frac{1}{\sqrt{2}} \left[ 1 - {qg_s\over 2\vert p\vert} +
o(g_s^3)\right]
\ .
\end{equation}
Expanding out the expression for the Casimir energy, and
using the special values  of the 
complete elliptic integrals at $k=\frac{1}{\sqrt{2}}$, one finds:
\begin{equation}
\label{expand}
{\cal E}_{\rm ren}\; =\; 
- {2  \pi^2 \over \Gamma({1\over 4})^4 L R}\;
{pQ_5\over  g_s} \  - \ 
 {q Q_5\over 4LR} \  +\  o(g_s)\ .
\end{equation} 

The leading term should be compared 
 to the holographic Wilson loop
computation in four-dimensional Yang-Mills theory. 
With our conventions of measuring the energy, the result for
the quark/antiquark potential is  \cite{qq1,qq2}
\begin{equation}
{\cal E}_{q\bar q} = -
 { 2 \pi^2 \sqrt{4\pi g_{YM}^2 N}\over \Gamma({1\over 4})^4 LR }
\end{equation}
The two results are identical if one notes that the
 radius of AdS5  is given (in string units) by $L^4= 4\pi
g^2_{YM}N$, whereas for AdS3 it is determined by $L^2=Q_5$. This is
no surprise since both calculations
minimize a pure tensive  energy,  which is proportional to
the geometric length of the string. From the sigma-model point of
view this Casimir energy is harder to understand, as we will
explain in the next  subsection. 
Note,  finally,   that the second term in the expansion \eqref{expand}
looks like a renormalized contribution to
the central charge.


\subsection{Symmetric product orbifolds and moduli flows}

We will now consider this computation from the point of view of the
space-time CFT.  Before taking the near-horizon limit,  the
configuration is described by a string junction built out of the
strings with charges $\vec q_1$, $\vec q_2$ and $\vec q_3 = \vec q_1 +
\vec q_2$. We assume that the string  $\vec q_1$, and therefore also
the string
$\vec q_3$, are much heavier than the string $\vec q_2$. Geometrically this
implies that the ``probe string'' $\vec q_2$ is perpendicular to both
$\vec q_1$ and $\vec q_3$,  which are parallel.

We now take the usual AdS/CFT decoupling limit. From the bulk point of
view we obtain the supergravity configuration of the previous section,
where the light string $\vec q_2$ is treated as a probe brane. From
the boundary point of view the two heavy strings $\vec q_1$ and $\vec
q_3$ each flow to a conformal field theory in the infrared. The two
conformal field theories are glued together along the string junction.

What is the fate of the string $\vec q_2$? Since we take the
near-horizon limit in the direction perpendicular to the heavy
strings, in this approximation there is no non-trivial decoupling
limit of the light string $\vec q_2$. Its worldsheet excitations are
in the perpendicular direction. Therefore in the IR limit holography
dictates that  the zero modes of the $\vec q_2$ string
survive as moduli of the space-time CFT.
 There are no separate degrees
of freedom living at the intersection point of the string
junction. The junction is basically a junction of $(p,q)$ strings in
the background of fivebranes. A junction of $(p,q)$ strings can be
thought of as a single M-theory M2 brane wrapping a suitable one cycle
of a two-torus. Since this is a smooth membrane configuration 
 there should be  no localized degrees of freedom at the
intersection. Thus the final space-time theory consists of two CFT's
on a half cylinder, separated by domain walls of the type we have been
discussing so far.

It remains to discuss the way in which the two CFTs are glued together
along the defect line. By general principle the CFT labeled by the
charge $\vec q$ can be identified with a ${\cal N}=(4,4)$ sigma model
with target space $M_N$. Here $M_N$ is a hyper-K{\"a}hler manifold that
is a deformation of the symmetric product $S^NM =M^N/S_N$ with
$N={\vec q}^2/2$ for $M=T^4$ and $N={\vec q}^2/2 + 1$ for $M=K3$. In
general this deformation is determined by the charge vector $\vec q$
and the original moduli of the string theory background. It will
include both metric deformations and sigma model $B$-fields. The
metric deformations will include turning on twist fields in the
orbifold description of the symmetric product. The CFT $B$-fields
correspond to space-time RR backgrounds.

The naive supergravity dual will have $B=0$ and will be strongly
coupled, in the sense of both small target space volume and large
twist field hyper-K{\"a}hler deformations. $T$-dualities do not in general
suffice to relate small volume to large volume sigma models.
The weakly coupled space-time CFT --- the analogue of perturbative
Yang-Mills theory in four dimension --- is given by the orbifold CFT
on $S^NM$ at large volume. In this regime the supergravity becomes a
string theory with large RR fields (since $B_{CFT} = 1/2$) and large
(ten-dimensional) string coupling constant.

Therefore, as always, the supergravity and weak-coupling CFT
computations are in disjunct regimes. We will see that they indeed
give qualitatively different behaviour for the domain walls. This is
to be expected in view of the four-dimensional Wilson loop computations,
where one observes a similar discrepancy. Alternatively, notice that a
weakly-coupled CFT should have operators of arbitrary spin in its
spectrum, and hence cannot be described by pure  supergravity.

In the case of general charge vectors $\vec q_1$ and $\vec q_3=\vec
q_1+ \vec q_2$ the two CFTs will have different central charges and
will be described by sigma models with target spaces of different
topology $M_{N_1}$ and $M_{N_3}$. In a semi-classical regime the
gluing of the two sigma models will be given by a D-brane $Y \subset
M_{N_1} \times M_{N_3}$. (Locally such a brane can be given by the
graph of a function $\varphi:\ M_{N_1} \times M_{N_3}$. Globally, we
are dealing with a generalized function, know mathematically as
a correspondence.)

We will simplify now the discussion to the case where the emitted
string is either a pure fundamental string or a pure D-string. In both
cases we will compare the CFT with the supergravity computation.

\subsubsection{Fundamental strings}

Let us start with the case where the string $\vec q_2$ is a
fundamental string.  In this case we are always dealing with a bound
state of $Q_5$ NS 5-branes and $Q_1$ fundamental strings. This system
is dual to the famous D1-D5 system that has been studied
extensively. In this case the space-time CFT is well-known. It is
given by a sigma model on the target space ${\cal M}_{Q_5,Q_1}$ ---
the moduli space of charge $Q_1$ instantons in a $U(Q_5)$ Yang-Mills
theory on the four-manifold $M^4=T^4,K3$. For relative prime
$(Q_1,Q_5)$ this moduli space is indeed a hyper-K{\"a}hler deformation
of the symmetric product $S^NM$ with $N=Q_1Q_5$.

We will be considering a string junction with $\vec q_1=(Q_1,Q_5)$,
$\vec q_2 = (q,0)$ and $\vec q_3=(Q_1+q,Q_5)$. Physically the process
whereby $q$ fundamental strings are absorbed corresponds to addition
of $q$ extra pointlike instantons in the Yang-Mills theory. The gluing
map
\begin{equation}
\varphi:\  {\cal M}_{Q_5,Q_1} \to {\cal M}_{Q_5,Q_1+q}
\end{equation}
can be described informally as follows. Place $q$ coincident pointlike
instantons at a point $x$ in the four-manifold $M$ and add this
solution to the smooth $Q_1$-instanton. This map depends on the choice
of point $x \in M$. The map $\varphi$ gives an isometric embedding of
${\cal M}_{Q_5,Q_1}$ into ${\cal M}_{Q_5,Q_1+q}$. This can be easily
seen in a local computation of instantons on ${\bf R}^4$ where the
ADHM construction can be used. We will give a more precise analysis of
the geometry in section 4.3.3.

The corresponding D-brane that describes the gluing with the use of
the folding construction is now given geometrically by the graph of
the map $\varphi$ in the product of the two instanton moduli
spaces. It has dimension $4Q_1Q_5$.

Let us sketch now an argument why the
 Casimir energy of two such domain walls should  be 
straightforward to compute
 in this regime. We have Neumann boundary conditions
for $4Q_1Q_5$ bosons and fermions. The remaining $4qQ_5$ bosons and
fermions that describe the normal directions to ${\cal M}_{Q_5,Q_1}$
will have Dirichlet boundary conditions. Because of the isometric
embedding there will be no jump in the CFT moduli, once we have
canonically normalized the kinetic terms in the sigma models. So in
this case the sole contribution to the Casimir energy will be the jump
in the central charge
\begin{equation}
\Delta c = 6 q Q_5.
\end{equation}
If we separate the two domain walls over a distance $2 \ell$,  this gives
a Casimir energy
\begin{equation}
{\cal E} = - {\pi \over 48 \ell} \Delta c=
 - {\pi \over 8 \ell} q Q_5
\end{equation}
This answer coincides with the supergravity computation
(\ref{q-casimir}) if we use $\ell= LR$ for the domain size.

\subsubsection{D-strings}

Let us now concentrate on the other limit where the absorbed string is
a pure D-string with charge $p$. In this case the interpretation in
terms of instanton moduli spaces is less clear. If we dualize the NS
5-brane to a D5-brane to obtain a gauge theory formulation, the
addition of a D-string is equivalent to adding a fundamental string. This
is represented by an electric flux tube in the gauge-theory instanton
background. From the gauge-theory point of view this description of
the CFT limit is not well understood.

In this case the string charge
vectors $\vec q_1$ and $\vec q_3$ will satisfy
\begin{equation}
{\vec q_1}{}^2={\vec q_3}{}^2
\end{equation}
Therefore the two sigma models have equal central charge and are in
fact topologically isomorphic. Both are given by a deformation of the
symmetric product $S^NM$. They only differ in the value of the
deformation moduli. One way to understand this is that there is a
U-duality transformation $U \in O(5,5+n;{\bf Z})$ that maps $\vec q_1$
to $\vec q_3$
\begin{equation}
U(\vec q_1) = \vec q_3.
\end{equation}
By definition the transformation $U$ does not leave the charge vector
$\vec q_1$ invariant.  Therefore it does not descend to a T-duality of
the sigma model.

We can understand this change in the moduli as follows. We
start with a string with charge $\vec q_1$. In the IR limit the moduli
of the CFT are obtained from the moduli of the string theory
background through the attractor formalism. That is, the scalars flow
towards their fixed values at the horizon where they satisfy $\vec
q_{1,R}=0$. 

Abstractly, if ${\cal N}$ represents the full string theory moduli
space, then ${\cal N}$ contains a sublocus ${\cal N}_{\vec q_1}$ that
represent the fixed scalars for the charge vector $\vec q_1$. The
moment we absorb the D-string, the charge vector changes to $\vec
q_3=\vec q_1 +\vec q_2$, and no longer satisfies the fixed scalar
condition. The moduli will now start to run along the attractor flow
lines to the new fixed point locus ${\cal N}_{\vec q_3}$
where $\vec q_{3,R}=0$.

Note that the U-duality transformation $U$ will map the fixed point
locus ${\cal N}_{\vec q_1}$ to the fixed point locus ${\cal N}_{\vec
q_3}$. We can therefore globally compare the values of the moduli of
the two spacetime CFTs.

The flow in the moduli can be computed exactly using for instance the
formalism developed by Mikhailov \cite{Mikhailov:1999fd}. 
Here we just mention the first order
effect in the D-brane charge $p$. The leading flow in the moduli is a
contribution to the RR 0-form and 4-form fields.which are of the form
\begin{equation}
\delta(\hbox{RR-moduli}) \sim {p\over g_s Q_1} .
\end{equation}
At the symmetric product point we would expect that we can use a free
field theory computation with order $Q_1Q_5$ free fields. 
To get an idea what the answer will look like we can first do
the calculation in case the 
target space is an $n$-dimensional torus with
constant metric $G_{\mu \nu}$ and $B$-field $B_{\mu \nu}$.
If we normalize $B$ in such a way that it has integral
periods, and assume that the domain wall separates two CFT's
with equal metric, but with $B$-fields $B$ and $B+\delta B$, 
the Casimir energy is proportional to
\begin{equation} \label{pq01}
{\cal E} \sim -\frac{1}{L} G^{\alpha\beta} G^{\gamma\delta}
 \delta B_{\alpha\gamma} \delta B_{\beta\delta}.
\end{equation}
If we apply the same equation to the symmetric product CFT,
we obtain qualitatively the following result.
Since the volume of $S^N M$ is of the form $f(Q_1,Q_5)V^N$, 
the Casimir energy must be of the form
\begin{equation} \label{pq02}
{\cal E} \sim -\frac{1}{L} \frac{g(Q_1,Q_5) }{V} 
\left({p\over g_s Q_1}\right)^2.
\end{equation}
The fact that the $B$-field in question is dual to a 
two-cycle with self-intersection $Q_1 Q_5$ suggests
that $g(Q_1,Q_5)$ is proportional to $Q_1 Q_5$
up to a factor of order unity, but a more careful
analysis is required to make this precise. In any case,
there will never be a precise agreement between the supergravity
answer and the CFT calculation, because the first one
is proportional to $p$, and the second one is proportional 
to $p^2$. This mismatch is similar to the disagreement found
in $N=4$ SYM, where the supergravity answer is proportional
to $\sqrt{g^2_{YM} N}$, whereas the answer at weak coupling
obtained in the gauge theory is proportional to
$g^2_{YM} N$.

\subsubsection{Domain walls and Nakajima algebras}

In the case of where the AdS string is a pure fundamental string
there is an interesting mathematical analogy. Namely, the string
junctions made completey out of fivebranes and strings have an elegant
interpretation in terms of the correspondences studied in
\cite{Grojnowksi,Nakajima1,Nakajima2}. In fact we can even consider a more
general situation where three strings join with charges $\vec q_1,\vec
q_2,\vec q_3=\vec q_1 + \vec q_2$, and where each string is built out
of fivebranes and fundamental strings, i.e. the charge vectors are of
the form $\vec q_i = (Q_5,Q_1)$. The folding construction that we have
used previously to describe the junction of two CFTs can easily be
extended to describe a junction of more than two CFTs. In that case
the junction conditions are given in terms of a boundary state in the
multiple tensor product of the corresponding Hilbert spaces. For sigma
models that translates into a D-brane in the cartesian product of the
target spaces.

For example, in the case of a three-string junction the domain wall
will correpond to a boundary state in
\begin{equation}
|B \rangle\rangle \in {\cal H}_1 \otimes {\cal H}_2 \otimes {\cal H}_3.
\end{equation}
Each of the three world-sheet theories will flow in the IR to a sigma
model with as target the instanton moduli space ${\cal M}_i$,
$i=1,2,3$. The junction is therefore geometrically, at large volume,
given by a brane $Y$ in the direct product
\begin{equation}
Y \subset {\cal M}_1 \times {\cal M}_2 \times {\cal M}_3^*
\label{cor}
\end{equation}
Here the asterix on ${\cal M}_3$ indicates that we choose minus the
holomorphic symplectic form. 

The ``correspondence'' $Y$ has an elegant mathematical interpretation
when we represent the instanton parametrized by ${\cal M}_i$ in terms
of a holomorphic vector bundle or more general a coherent torsion free
sheaf ${\cal E}_i$. The locus $Y$ is then given by triples $({\cal
E}_1,{\cal E}_2,{\cal E}_3)$ that are related by an exact sequence
\begin{equation}
0 \to {\cal E}_2 \to {\cal E}_3 \to {\cal E}_1 \to 0.
\end{equation}
That is, the sheaf ${\cal E}_3$ can be obtained by an extension of
${\cal E}_1$ by ${\cal E}_2$ (or vice versa taking duals). The locus
$Y$ is a complex Lagrangian submanifold in the given complex
symplectic form.  It therefore corresponds to a D-brane that preserves
the diagonal ${\cal N}= 4$ superconformal algebra in the tensor
product CFT.

Note that more generally the D-brane $Y$ is classified by an element
of the K-theory group associated to the product ${\cal M}_1 \times 
{\cal M}_2 \times {\cal M}_3^*$. Domain walls in the 1-brane/5-brane
system are therefore a natural place where the K-theory of instanton
moduli spaces occurs  within string theory.

By very general arguments, a correspondence of the form (\ref{cor})
naturally leads to linear maps on the level of the cohomology of the
moduli spaces ${\cal M}_i$. More precisely it gives rise to a map
\begin{equation}
\varphi:\ H^*({\cal M}_1) \times H^*({\cal M}_2) \to H^*({\cal M}_3)
\end{equation}
This map given by pull-back of a differential form on ${\cal M}_1
\times {\cal M}_2$ to the triple product, followed by restriction to
the D-brane locus $Y$ and push-forward (integrating over the fiber) to
${\cal M}_3$. In a formula 
\begin{equation}
\varphi(\alpha,\beta) = \pi_{3,*}\left(\pi_1^*\alpha \wedge \pi_2^* \beta
\cdot [Y]\right)
\end{equation}
The adjoint is given by following this series of maps in the other
direction.

If $\vec q_2=(0,n)$ is built only out of strings and  no fivebranes, the
moduli space ${\cal M}_2$ parametrizes skyscraper sheaves and is
simply given by a copy of the four-manifold $M$.  Therefore for every
element $\alpha \in H^*(M)$ the map $\varphi$ defined above reduces to
a map
\begin{equation}
\alpha_n : H^*({\cal M}_{1}) \to H^*({\cal M}_{3})
\end{equation}
with
\begin{equation}
{\cal M}_1 = {\cal M}_{Q_5,Q_1},\qquad {\cal M}_3 = {\cal
M}_{Q_5,Q_1+n}
\end{equation}
Its adjoint $\alpha_{-n}$ is defined similarly. These maps have been
studied extensively in the mathematical literature.  In particular for
the case $M={\bf C}^2$ Nakajima \cite{Nakajima2} has shown that the
operators $\alpha_n$ give rise to a Heisenberg algebra
\begin{equation}
[\alpha_n,\beta_m]=n \delta_{n+m,0} \int_M \alpha \wedge \beta
\label{heisenberg}
\end{equation}

We want to remark that these maps get a natural interpretation in the
context of CFT domain walls that are the subject of this
paper. Consider such a domain wall on the cylinder in the closed
string channel labeled by some index $I$. That is, consider the domain
wall along a space-like slice where it is interpreted as a Euclidean
instantaneous brane. Since this brane connects $CFT_1$ and $CFT_3$, it
gives rise to a map on the level of Hilbert spaces
\begin{equation}
\varphi_I:\ {\cal H}_1 \to {\cal H}_3.
\end{equation}
(This map will strictly speaking not exist at the level of proper
Hilbert spaces since it will map normalizable states to unnormalizable
states.) If we restrict the map $\varphi_I$ to ground states we expect
to find a generalization of  Nakajima's map. This suggest that there is
an interesting exchange algebra of such domain walls that should give
rise to the commutation relations (\ref{heisenberg}) in the quantum
mechanics approximation. These relations have also been studied by
Harvey and Moore in the context of the algebra of BPS states in
\cite{Harvey-Moore}. It would be very interesting to connect these
 two points of view more directly.

To be concrete,  we  give the  expression in the  much simpler
case of  the free-field domain wall that we studied in
section 2. For a given  $\vartheta$  the corresponding operator
\begin{equation}
S_\vartheta : \ {\cal H} \to {\cal H},
\end{equation}
with ${\cal H}$ the free-boson Fock space, is given by (in the same
canonical normalization as  in section 2)
\begin{equation}
S_\vartheta = \prod_{n>0} \exp\left(-\cos 2\vartheta\; a_n \bar a_n\right)
\prod_{n>0} (\sin 2 \vartheta)^{a_na_{-n} + \bar a_n \bar a_{-n}}
\prod_{n>0} \exp \left(\cos 2\vartheta\; a_{-n} \bar a_{-n}\right)
\end{equation}
One easily verifies that for $\vartheta
 ={\pi\over 4} ,{\pi\over 2}$ (that is
$\lambda =1, \infty$) this gives the correct expresion for a
completely permeable,  or completely  reflective wall
\begin{equation}
S_{{\pi\over 4}} = {\bf 1},\qquad S_{\pi\over 2} = |0\rangle\langle 0|.
\end{equation}

\boldmath
\section{Outlook}
\unboldmath

  An  interesting problem  for future work is to 
construct explicit models of  permeable interfaces between  
strongly-coupled conformal field theories. 
As explained in section 3.2, one needs to 
find boundary states of  tensor-product theories, which cannot be
expressed in terms of  Ishibashi states of the individual  factors.
 One could try, for example, to embed   WZW D-branes as
`non-factorizable'  states in the  $G/H\otimes H$ theory. 
Another place where to look for such defects
 is in the product of two WZW models with different Kac-Moody
levels,
for which K-theoretic arguments
predict   more charges
than those that can be  accounted for by
elementary WZW D-branes.\footnote{We thank
Volker Schomerus for pointing  out these arguments.}
Besides their intrinsic mathematical interest, such examples,  if they
exist,  could  find 
applications in condensed-matter physics.

Another natural  question raised by our work is
whether one can  construct 
a string theory whose worldsheet contains
permeable defects. One immediate difficulty with this idea
is that if both CFT1 and CFT2  contain
time coordinates in their target spaces,  
we need two Virasoro symmetries in order to 
remove all the negative-norm states from the spectrum.
But the generic 
permeable walls only preserve one symmetry, as we have explained.
One can  try to circumvent this problem by asking,  
say,  that CFT2  have a  Euclidean target space. 
The no-ghost theorem
requires, however, in this case
that  the total central charge of  CFT1$\otimes$CFT2  be 26.
Thus, in the product theory one has a single  time coordinate,
a  central charge  26, and a conventional
 conformal boundary state.
This looks like
 a standard open-string theory on a regular D-brane!  Whether there
could be loopholes in the above argument is a question that deserves
further thought.

\vskip 0.3cm

{\bf Acknowledgments}: We thank  the ITP  at
Santa Barbara  and  the organizers of the `M-theory' program  for their
kind hospitality during  the initial stages of this work. 
We are also grateful to  C. Albertsson, C. Callan,
O. DeWolfe, D. Freedman, J. Fr{\"o}hlich, D. Kutasov, U. Lindstr{\"o}m, 
A. Ludwig, J. Maldacena, E. Martinec, G. Moore,  N. Reed, 
R. Russo, V. Schomerus, J. Schwarz,  E. Verlinde and M. Zabzine
 for useful conversations.
This research  is  supported in part by the European networks 
``Superstring
Theory'' HPRN-CT-2000-00122 and ``The Quantum Structure of Spacetime''
HPRN-CT-2000-00131, and by DOE grant DE-AC03-76SF00098. 

\vskip 0.6cm


\noindent {\bf\large
Appendix A: calculation of Casimir energy}

\vskip 0.3cm

 In this appendix we calculate the Casimir energy for the set up
described in section 2.2.  We consider a real, massless  
scalar  field $\tilde \phi$ in the interval
$[-L, L\; ]\ $ , with Dirichlet boundary conditions
 at the endpoints. This is a choice of convenience  that does not affect
our  final result.  The  action of $\tilde \phi$ is rescaled inside the
subinterval $[-d , d \; ]\ $  , where  $d<L$. This 
rescaling amounts to a change in radius, 
as discussed in the main text.
The general plane-wave solution is of the form:
$$
\phi(x,t) =  e^{i\omega t}\times \; 
\Biggl\{ 
\begin{array}{lll}
A_1 \;{\rm sin}(\omega x +\delta_1)\ \ \ \ \ {\rm for} 
 &\quad & \   x\in  [-L \;, -d \;] \cr
A_2 \;{\rm sin}(\omega x +\delta_2)\ \ \ \ \ {\rm for}
  &\quad & \   x\in  [-d\;, \; d\;] \cr
A_3 \;{\rm sin}(\omega x +\delta_3)\ \ \ \ \ {\rm for}
  &\quad & \   x\in \; [\;d\;, \;L\;] \cr  
\end{array}
\eqno(A1)
$$
The Dirichlet boundary conditions at $x=\pm L$ imply:
$$
\delta_1 = \omega L\;  ({\rm mod}\pi)\ \ \ \ {\rm and}\ \ \ \ 
\delta_3 = -\omega L\; ({\rm mod}\pi)\; .
\eqno(A2)
$$
The gluing conditions \eqref{bc}  at the two domain walls, on the other hand,
read:
$$ 
{\rm tan}(-\omega d  +\delta_1)\;  = \; \lambda^2\; 
 {\rm tan}(-\omega d   +\delta_2)\ ,
\eqno(A3)
$$
and
$$
{\rm tan}(\;\omega d  +\delta_3)\;  =\;  \lambda^2\; 
 {\rm tan}(\;\omega d   +\delta_2)\ .
\eqno(A4)
$$
Putting together  (A.2), (A.3) and (A.4)  leads to a transcendental
  equation for the 
allowed frequencies, 
$$ 
{\rm tan}\left[\;\omega (d-L)\right] \; = \;  \lambda^2 \;
 {\rm tan}(\omega d +\delta_2)\ \ \ 
{\rm with} \ \ \  \delta_2 = 0\; {\rm mod}\left(\frac{\pi}{2}\right)\ .
\eqno(A5)
$$

 We can solve this equation analytically 
in the limit $L\to\infty$ with $d$  held fixed. Let us write
$$
\omega_n \equiv  {\pi \over 2 L}\; (n -\epsilon_n)\ .
$$
The `unperturbed' spectrum, in the absence of walls,  has
$\epsilon_n=0$. Assuming  that $\epsilon_n$  stays  bounded,
so that 
we can neglect terms of  $o(\epsilon_n/L)$ in the equation, we  find:
$$
\epsilon_n \; = \; 
\Biggl\{ 
\begin{array}{lll}
\; ({2/\pi})\;{\rm arctan}\left[ \lambda^2 \;
 {\rm tan}({n\pi d/ 2L})\right]
- {nd/ L} 
 \ \ ({\rm mod}\;2) \ \  &\quad & \  {\rm for}\ $n$ \ \ {\rm even} \ ,\cr
\ \ \ \  &\quad & \  \cr 
\; ({2/ \pi})\;{\rm arctan}\left[ \lambda^{-2} \;
 {\rm tan} ({n\pi d/ 2L})\right] - {nd/ L} 
 \ \ ({\rm mod}\;2) \ \  &\quad & \  {\rm for}\ $n$ \ \ {\rm odd} \ .\cr
\end{array}
\eqno(A6)
$$
We have chosen  $\delta_2 = 0$ for even $n$,  and
$\delta_2 = \pi/2$ for odd $n$, so that $\epsilon_n$ vanishes
when $\lambda =1$ (no walls).
We will also 
choose the branch of the arctangent
such that  $-1\leq \epsilon_n \leq 1$. This ensures that 
$\omega_n$ is closest 
 to its `unperturbed' value (and  is consistent with our
assumption of bounded  $\epsilon_n$).

Since  $\omega_n$ and $\omega_{-n}$
 correspond to the same wavefunction,  the
Casimir energy reads:
$$
{\cal E} \;  = \; \sum_{n=1}^\infty\;  {1\over 2} \omega_n\ .
\eqno(A7)
$$
  The result  is of course
UV divergent, so we must perform the summation
 with great   care.
We will use the standard regularized formula:
$$
\sum_{n=1}^\infty (n- \alpha) = 
-{1\over 12} + \frac{1}{2}\alpha (1-\alpha)\ .
\eqno(A8)
$$
The trick is to pick  $L=Nd/2$   for 
integer $N$ (which we will send  eventually
to infinity). If the limit $L\to\infty$ exists
it should not matter how we approach it. 
With this choice  the frequency shifts are periodic~:
$$
\epsilon_n = \epsilon_{n+N}\ \ , 
\eqno(A9)
$$
Expressing the arbitrary  positive  integer $n$  as follows~: 
 $n = lN-k$ with $l=1,\cdots , \infty$ and 
$k = 0, 1,  \cdots , N-1$, we 
decompose  the Casimir energy into  $N$ 
sums regularized separately as in 
 (A8)\;.\footnote{The reader can be reassured about this manipulation of
divergent sums  by checking, for instance,  that the formal
identity $\sum_{1}^\infty n
= \sum_{k=0}^{N-1}  \sum_{l=1}^\infty  N (l - \frac{k}{N}) $
 stays  valid after
regularization of  the $n$- and $l$-sums as in  (A8).
}
The result after some algebraic rearrangements is
$$
{\cal E} = -\frac{\pi}{48L} -  \frac{\pi}{4d}\;\sum_{k=0}^{N-1}
 \frac{\epsilon_k}{N} \left(1 - \frac{2k}{N}-
 \frac{\epsilon_k}{N}\right)\ .
\eqno(A10)
$$

The first term is the `unperturbed' Casimir energy,  which vanishes in the
$L\sim N \to\infty$ limit. The
 term quadratic in $\epsilon_n$ is also subleading, so we may  drop it in
this limit, as well.
 For $0\leq k < N/2$,
the shift $\epsilon_k$ is in the desired range 
(between $-1$ and $1$)   and we can
perform the remaining sum in (A10) as it stands.
The other half-range,  $N/2\leq k < N$, contributes an equal
amount to the energy, as can be seen by changing variable to
$\tilde k = N-k$, and using  the fact that $\epsilon_{N-\tilde k}=
- \epsilon_{\tilde k}$. 

Defining finally the continuous variable $2y= {\pi}( 1 - 2k/N)$,
and using standard  trigonometric identities, 
 leads   to the integral  expression for the Casimir energy:
$$
{\cal E}
 = \frac{1}{\pi^2 d} \int_0^{\pi/2}  dy\; y \;
\left[\;  2 y -  
\;{\rm arctan}(\lambda^2 {\rm tan}y) -
\;{\rm arctan}(\lambda^{-2} {\rm tan}y) 
\right]
\ .
\eqno(A11)
$$
This formula passes several consistency checks~:  it vanishes
for $\lambda =1$,  it has manifest symmetry under inversion of
$\lambda$ (which is equivalent to a T-duality transformation),
 and  it gives the expected Casimir energy, 
${\cal E}=-\pi/48 d$ , in the case
of  perfectly-reflecting walls ($\lambda =0$).

  We can perform the integral (A11) explicitly by using the dilogarithm
function ${\rm Li}_2(z)$.  This has the series and integral representations
(for $z<1$) 
$$
{\rm Li}_2(z) = 
\sum_{1}^\infty {z^m\over m^2} = \int_0^z {{\rm log}(1-w)\over w}\; dw\ .
\eqno(A12)
$$
Many of its properties can be found in ref. \cite{lewin}.
It obeys, in particular, the identity
$$
{\rm Li}_2(z) + {\rm Li}_2(-z) = {1\over 2}{\rm Li}_2(z^2)\ . 
\eqno(A13)
$$
We also  need the
integration formula \cite{lewin}
$$
\int_0^{\pi/2} {y^2 dy\over 1- P {\rm cos}(2y)} = {1+p^2\over 1-p^2}\;
\left[ \frac{\pi^3}{24} + \frac{\pi}{2}{\rm Li}_2(-p)\right] 
\eqno(A14)
$$
where 
$$
P = {2p\over 1+p^2}\ , \ \ \ {\rm with} \ \ \ p^2< 1\  .
\eqno(A15) 
$$
Integrating the right-hand-side of (A11) by parts,  and using the above
equations, puts the Casimir energy
 in the  compact form quoted in the main text:
$$
{\cal E} =  - \frac{1}{8\pi d }\;  {\rm Li}_2({\cal R}^2)\ \ 
\ {\rm with} \ \ \ {\cal R}= {1-\lambda^2\over 1+\lambda^2}\ . 
\eqno(A16)
$$
Here ${\cal R}$ is the reflection coefficient. For total reflection 
${\cal R}=\pm 1$,   and since   ${\rm Li}_2(1) =  \pi^2/6$, we find indeed
the standard Casimir energy of a scalar
 field. For weak reflection, the energy
vanishes quadratically: ${\cal E} \simeq  - {{\cal R}^2}/8\pi d \; $ .

 The dilogarithm function has appeared in
 CFT and integrable models, in various
contexts (see for example \cite{NRT}). 
The  above interpretation as free-field Casimir energy is,
to the best of our knowledge, new.


\end{document}